\newcommand{\paperversion}{arxiv}
\newif\ifreentrancymotivation
\reentrancymotivationtrue

\providecommand{\paperversion}{draft}
\input{paperversions} 

\ifshowlinenums
  
\else
  
\fi

\ifblinded
  
\else
  
\fi

\ifcopyrightspace
  
\else
  
\fi

\ifdefstring{\paperversion}{arxiv}{%
  \documentclass[acmsmall,nonacm,review=false]{acmart}
}{%
  \documentclass[acmsmall,review]{acmart}
}
\ifshowpagenumbers
 \ifdefined
 \fi
\fi
\ifblinded
 \ifdefined
 \fi
\fi

\usepackage[utf8]{inputenc}

\iffinalformat
  \usepackage{pbalance}
\fi

\usepackage{booktabs}   

\usepackage{tikz}
\usetikzlibrary{calc}
\usepackage{wrapfig}
\usepackage{hyperref}
\usepackage{amsmath}
\usepackage{amssymb}
\usepackage{amsthm}
\usepackage{mathtools}
  \mathtoolsset{mathic}
\usepackage{mathpartir}
\usepackage{pl-syntax}
\usepackage{xcolor}
\usepackage{xspace}
\usepackage{subcaption}
\usepackage[pages=20]{checkpagelimit}

\usepackage[small]{ttquot}

\ifshowcomments
    \usepackage[checkNoComments,margin]{aplcomments}
\else
    \usepackage[disabled]{aplcomments}
\fi
\newcommenter{todo}{1.0,0.2,0.2}
\newcommenter{GC}{0.0,0.5,0.0}

\usepackage{defs}
\usepackage{concrete-syntax}
\usepackage{functions}
\usepackage{information-flow}
\usepackage{meta-variables}
\usepackage{type-confusion}

\usepackage{cleveref}
  \crefformat{section}{\S#2#1#3}

\usepackage{url}
\usepackage[bibliography=common]{apxproof}
\usepackage{overrides}

\newcommand\cornell{
  \affiliation{%
    \institution{Cornell University}
    \city{Ithaca}
    \state{NY}
    \postcode{14850}
    \country{USA}
  }
}

\ifblinded
  \author{Anonymous authors}
\else
    \ifdefined\acmYear

        \author{Silei Ren}
        \authornote{Equal Contribution.}
        \affiliation{%
            \institution{Amazon}
            \city{Arlington}
            \state{VA}
            \postcode{}
            \country{USA}
        }
        \email{sr2262@cornell.edu}

        \author{Hanxi Chen}
        \authornotemark[1]
        \cornell
        \email{hc2296@cornell.edu}

        \author{Suraaj Kanniwadi}
        \cornell
        \email{suraaj@cs.cornell.edu}

        \author{Andrew C. Myers}
        \orcid{0000-0001-5819-7588}
        \cornell
        \email{andru@cs.cornell.edu}
    \else
        \author{Silei Ren, Suraaj Kanniwadi, Hanxi Chen, Andrew C.
        Myers \\ Amazon and Cornell University}
    \fi
\fi

\title{Type-Confusion Security}

\ifshowpagenumbers
\fi

\draftwarning

\begin{document}

\begin{abstract}
  Type systems are a lightweight way to obtain formal assurance that
code upholds security properties.  However,
decentralized systems, such as smart contracts and Model Context
Protocol ecosystems, have limited ability to constrain the
behavior of untrusted code.
In such systems, attackers may attempt to confuse trusted code by
supplying values that do not behave according to their advertised types,
an attack that we call \emph{type confusion}.
Well-known attacks such as confused deputy
attacks, reentrancy attacks, and control-flow hijacking
can be understood as instances of type confusion.
We define \emph{type confusion security} as the property that
unrestricted attackers who need not follow the type system have
no more power than limited attackers who fully obey it.
We formally define type confusion security as a real--ideal emulation,
and in the context of an information-flow-typed core calculus that
statically enforces security against well-typed attackers, prove that
its run-time checks also enforce security against ill-typed attackers.
In this system, programmers can write code while reasoning about
security only against well-typed attackers; the security guarantees
enforced by the type system still hold.  We mechanize a proof of type
confusion security in Lean.

\end{abstract}

\maketitle

\section{Introduction}

Static type systems are a powerful way to enforce security.
They allow programmers to reason about code in a high-level language
and rule out attacks before the code runs, often with
little or no run-time overhead.
In type systems for information flow control (IFC), _labeled_ types serve as
expressive security specifications for code: they describe confidentiality and
integrity requirements, identify contexts in which code may execute securely,
and also enable enforcement of properties like
noninterference~\cite{GM82,Schneider-EM}, reentrancy security\cite{serif21} and
nonmalleable information flow \cite{nmifc}.

However, purely static reasoning is fragile in systems where
adversaries may not follow the rules.
For example, smart contracts, cross-chain protocols, federated systems
like Bluesky or OAuth 2.0, and Model Context Protocol (MCP)
ecosystems routinely interact across trust boundaries.
In such settings, trusted code may receive ill-typed values from an
adversarial principal or invoke remote procedures whose
implementations are controlled by an adversary.
Our name for an adversary-induced mismatch between the expected and
actual types is \emph{type confusion}.

Depending on the type system, type confusion has different
security implications.
Its most obvious form involves unchecked casts
between ordinary types, which can violate memory
safety. Further, an integer can be used at run time
as a function pointer, leading to vulnerabilities such as _control-flow
hijacking_ \cite{cfi, mcfi}: calling the function jumps
to an invalid or unauthorized address.

Type confusion can be more insidious in security type
systems, where labels on types
express security requirements for authorization and information flow.
%
As we illustrate, \emph{confused-deputy attacks (CDAs)}~\cite{confdeputy} and
\emph{reentrancy attacks}~\cite{serif21} are examples of such phenomena.
Such attacks are not merely hypothetical: recent high-profile security
breaches in blockchain smart contracts have been attributed to CDAs and
reentrancy attacks~\cite{dexible-attack, polynetwork-attack,
lifi-attack, dao-hack-analysis} that exploit mismatched
expectations about the behavior of callback code. 
In both cases, attackers exploit type-confused function pointers to induce
insecure control flows.

Solutions for such type confusion are unsatisfactory. Existing techniques like
gradual typing~\cite{quest-complete, hlio} don't address adversarial settings:
every term is either dynamically or statically typed, the runtime is trusted,
and users don't actively misrepresent types.
State-of-the-art language-based IFC systems like
SCIF~\cite{scif-arxiv} and Fabric~\cite{fabric09} partially address
these attacks but lack formal guarantees.  Prior formal
works \cite{rajaniAccessControlCapabilities2016, serif21} show that
IFC, extended with run-time checks, can enforce security against CDAs
and reentrancy attacks independently, but not in a unified framework.

Our core contribution is a unified, extensional, and rigorous description of
security for the above attacks, as type confusion security. Further, we
construct effective dynamic mechanisms against type confusion, and highlight
their semantic principles. Our contributions are summarized below:

\paragraph{Type-confusion security}
We define \emph{type-confusion security} in the style of simulation-based
security~\cite{UC}: attackers gain no advantage by confusing types because such
attackers can be simulated by ones who do not.
%
%
%
Type-confusion security ensures robust hyperproperty preservation
(RHP)~\cite{abateB0HPT19, rhp-uc}, a compiler correctness property which
requires that whatever properties the type system enforces---for example,
memory safety or noninterference---hold even in the presence of real-world
attackers.
Thus, programmers may reason about security in the simpler, well-typed
ideal world, even though deployed code
interacts with real-world adversaries that do not respect the type system.




\paragraph{Provable security enforcement of type confusion}

We design dynamic mechanisms that provably enforce
type-confusion security for an information-flow-typed language.
It might seem that it suffices to simply perform a dynamic check that mirrors
its static counterpart. However, we show that run-time enforcement
must be carefully designed to ensure that the dynamic checks are
strong enough, performed _only_ by honest parties, and depend _only_
on trusted run-time information.

\paragraph{Provable enforcement of real-world security}
This paper defines novel, simple run-time checks that
protect against CDAs.
It also extends the authority-lock mechanism of
SeRIF~\cite{serif21} to enforce reentrancy security more flexibly.

\paragraph{Language model and formalization}

To explore type-confusion security and its enforcement mechanisms, we
define a core calculus of a decentralized system with remote procedure calls (RPC).
%
In the setting of this core calculus, we mechanize all of
our security results in Lean~\cite{moura2021lean}.
%

\smallskip
The rest of the paper is structured as follows.
\Cref{sec:motivating} introduces motivating examples and the key
intuition behind the run-time checks. \Cref{sec:formal-theorems}
provides a formal overview of type-confusion security. \Cref{sec:calculus} defines the ideal and
real calculi. \Cref{sec:security} proves the main simulation
theorem. \Cref{sec:applications} applies the theorem to prove
security against CDA and reentrancy attacks. \Cref{sec:related-work} discusses
related work and concludes.

\section{Background and Motivation}

\label{sec:motivating}

\begingroup
\lstset{
    keywordstyle=[2]\bfseries\color{black},
    keywordstyle=[4]\color[rgb]{0.08,0.32,0.24},
    morekeywords=[3]{exchange, oracle, alice, adv, me},
    morekeywords=[4]{swap, quote, pay, onReject, ignore},
    deletedelim=[s]{\{}{\}},
    moredelim=[s][\bfseries\color{darkBlue}]{\{U}{\}},
    moredelim=[s][\bfseries\color{darkBlue}]{\{T}{\}}
}

\subsection{Confused Deputy Attacks}

A confused-deputy attack
(CDA)~\cite{confdeputy} happens when an attacker tricks a
trusted party (the _deputy_) into misusing its authority
to effectively escalate the attacker's privileges.
\Cref{fig:cda} shows a simplified version of a CDA against a token exchange
service.  It captures the spirit of several notorious recent attacks causing
millions of dollars in losses~\cite{dexible-attack, polynetwork-attack,
lifi-attack}.

Here, the exchange service is the trusted confused deputy.
In an honest execution, the user calls the exchange contract with an
exchange request, along with the user's payment function
\lstinline{alice.pay()} for the exchange service to extract funds.
When called, the exchange contract queries a price oracle for the
exchange rate, then calls the user's payment function, and finally
pays the user.
However, an attacker might pass the exchange's own payment handle
\lstinline{exchange.pay()} instead, causing the exchange service to
pay the user (line 3) when it should receive payment from the user.

The classic solution to CDA is capabilities~\cite{confdeputy}, but a less error-prone
solution can be obtained by viewing CDA through the lens of secure
information flow, where it becomes clear that CDA is a form of type
confusion.

\label{sec:cda}
\begin{figure}[t]
    \centering
    \captionsetup{skip=4pt}
    \begingroup
\newcommand{\cdaMark}[1]{\tikz[remember picture,overlay]\coordinate (#1);}
\lstset{
    basicstyle=\scriptsize\ttfamily,
    numberstyle=\tiny\itshape,
    keywordstyle=[2]\bfseries\color{black},
    keywordstyle=[4]\color[rgb]{0.08,0.32,0.24},
    frame=none,
    aboveskip=0pt,
    belowskip=0pt,
    morekeywords=[3]{exchange, oracle, alice},
    morekeywords=[4]{swap, quote, pay, receiver},
    morestring=[b]",
    stringstyle=\color{codeColor},
    showstringspaces=false,
    deletedelim=[s]{\{}{\}},
    moredelim=[s][\bfseries\color{darkBlue}]{\{U}{\}},
    moredelim=[s][\bfseries\color{darkBlue}]{\{T}{\}}
}
\scriptsize
\begin{tikzpicture}[
    remember picture,
    >=latex,
    codebox/.style={
        draw=black!65,
        fill=black!4,
        line width=0.5pt,
        rounded corners=3pt,
        inner sep=3pt,
        outer sep=0pt
    },
    boxlabel/.style={
        font=\scriptsize\sffamily,
        fill=none,
        inner xsep=2pt,
        inner ysep=1pt,
        anchor=south west,
        yshift=1pt
    },
    callarrow/.style={->, semithick, rounded corners=3pt,
        shorten <=1pt, shorten >=1pt},
    honestcall/.style={callarrow, draw=darkGreen!65!black},
    attackcall/.style={callarrow, draw=red!70!black, dashed},
    callnumber/.style={circle, line width=0.6pt, solid, fill=white, inner sep=0.5pt,
        minimum size=8pt, font=\tiny\sffamily},
    honestnumber/.style={callnumber, draw=darkGreen!65!black, text=darkGreen!65!black},
    attacknumber/.style={callnumber, draw=red!70!black, text=red!70!black}
]
    \node[codebox, anchor=west] (cda-user) at (0, 0.76cm) {%
        \begin{minipage}{0.43\linewidth}
            \vspace*{2pt}
            \lstinputlisting{figs/code/CDA-exchange-usr-IFC.code}
        \end{minipage}%
    };
    \node[codebox, anchor=west] (cda-attacker) at (0, -1.02cm) {%
        \begin{minipage}{0.43\linewidth}
            \vspace*{2pt}
            \lstinputlisting{figs/code/CDA-exchange-adv-IFC.code}
        \end{minipage}%
    };
    \coordinate (cda-right) at (\linewidth, 0);
    \node[codebox, anchor=north east] (cda-compiler)
        at ([yshift={-\baselineskip-1pt}]cda-user.north -| cda-right) {%
        \begin{minipage}{0.49\linewidth}
            \vspace*{2pt}
            \lstinputlisting{figs/code/CDA-exchange-IFC.code}
        \end{minipage}%
    };

    \node[boxlabel] at ([xshift=5pt]cda-user.north west)
        {Honest user {\bfseries\color{darkBlue}\{U\}}};
    \node[boxlabel] at ([xshift=5pt]cda-attacker.north west)
        {Attacker {\bfseries\color{darkBlue}\{U\}}};
    \node[boxlabel] at ([xshift=5pt]cda-compiler.north west)
        {Exchange {\bfseries\color{darkBlue}\{T\}}};

    \draw[honestcall]
        ([yshift=2pt]cda-user-call)
        -- node[midway, honestnumber] (cda-call-one) {1}
        ([yshift=2pt]cda-swap);
    \draw[attackcall]
        ([yshift=2pt]cda-attacker-call)
        -- node[pos=0.35, attacknumber] {1}
        ([yshift=2pt]cda-swap);
    \draw[honestcall]
        ([yshift=2pt]cda-receiver)
        -- ([yshift=2pt]cda-alice-pay);
    \node[honestnumber]
        at ([yshift=2pt]cda-call-one.center |- cda-receiver) {2};
    \draw[attackcall]
        ([yshift=2pt]cda-receiver-end)
        -- node[midway, draw=red!70!black, solid, rounded corners=1pt,
            fill=black!4, text=red!70!black, inner xsep=2pt, inner ysep=1pt,
            font=\tiny\sffamily] {dynamic check}
        ([xshift=-6pt,yshift=2pt]cda-compiler.east |- cda-receiver-end)
        -- node[midway, attacknumber] {2}
        ([xshift=-6pt,yshift=2pt]cda-compiler.east |- cda-pay)
        -- ([yshift=2pt]cda-pay);
\end{tikzpicture}
\endgroup
    \caption{Confused deputy attack on an exchange smart contract.
    The solid green arrows represent a well-typed usage of the service.
    Dashed red arrows show the control caused by an ill-typed attacker.
    Run-time checks on line 1 of \lstinline{exchange} cannot be trusted
    because they involve an untrusted caller.
    On a CDA run, the dynamic check on line 3 happens fully within the
    trusted parties because both the deputy and the victim are trusted.
    }
    \label{fig:cda}
\end{figure}
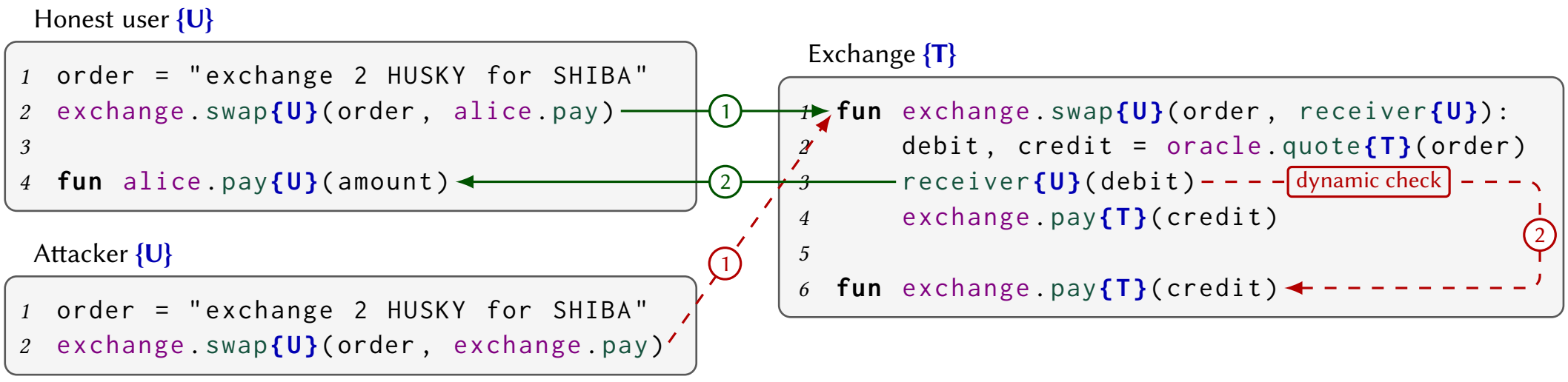

Assume a simple two-point lattice of information-flow labels
in which information is either labeled \lstinline"T" for 
``trusted'' or \lstinline"U" for ``untrusted''.
Function types
are annotated by labels; in particular, each function type includes
an _external pc (program counter) label_~\cite{serif21} which
controls the contexts from which it may be called, as seen
in \cref{fig:cda}.
A function labeled \lstinline{T} may only be called from trusted
contexts.  A
function labeled \lstinline{U} may be called from anywhere and passed
untrusted arguments, so a \lstinline{U}-function type is a subtype of
a \lstinline{T}-function type.
Although the exchange service is trusted to call any function in a
trusted context, we can design our IFC type system so that a call
through the untrusted variable \lstinline{receiver} (line 3) is
necessarily performed in an untrusted context \lstinline{U}.
%

If the static typing rules are enforced, the attack is prevented
because line 2 of the attacker code fails to type-check:
the attacker passes \lstinline{exchange.pay}, a \lstinline{T}-function pointer,
to a \lstinline{U}-labeled function parameter (\lstinline{receiver}).
Because function subtyping is contravariant on inputs, this argument
is rejected.
However, in the real world where attackers do not follow static typing
rules, run-time checks are needed to prevent CDAs.

Naively we might think to check the type of \lstinline{receiver} as
soon as \lstinline{exchange.swap} is called, but in an open
system we do not know this type and cannot trust the attacker to
provide it.  
Therefore, we perform run-time security checks lazily at function calls.
On line 3 of \lstinline{exchange.swap}, the exchange notifies the callee \lstinline{exchange.pay}
that it is being called from a \lstinline{U}-labeled context.
As the trusted callee \lstinline{exchange.pay} expects calls only from
\lstinline{T}-labeled contexts, it rejects the call at run time.

The obvious question is whether this simple dynamic check suffices
to prevent CDA.
We can show that it does by demonstrating
that with this check, the system is secure against type
confusion.


\subsection{Reentrancy Attacks}

CDAs are not the only attacks that could happen to the
exchange service of \cref{fig:cda}.
\Cref{fig:reentrancy} shows a \emph{reentrancy attack}
\cite{dao-hack-analysis}, in which the attacker causes the deputy to be
called again before it finishes its first call.
In this attack, the attacker writes a \lstinline{U}-labeled contract
that invokes \lstinline{exchange.swap} when called.
This callback causes the exchange service to be recursively
called many times before any call returns, adjusting
the exchange rate returned by the price oracle
in favor of the attacker.

As observed by Cecchetti et al.~\cite{serif21}, a reentrancy attack
exploits contracts where invariants over trusted state are expected
to hold before and after external function calls.
In the example from \cref{fig:reentrancy}, the (trusted) exchange rate
is expected to remain stable from line 2 to line 4.
The attacker indirectly influences that trusted state by diverting
the control flow to call \lstinline{oracle.quote()} multiple times
before \lstinline{exchange.pay()} is called.

Statically modeling reentrancy attacks requires reasoning about the
effects of functions on trusted state.
To account for that, we introduce a \textit{\bounding} on function
types, to upper-bound the level of effects during function execution.
The function \lstinline{exchange.pay()} is a
\lstinline|{T$\topbounded$T}|-function that can only be called in a trusted
context and updates the trusted state of the exchange. Bounding labels are
contravariant because a function with an untrusted \bounding also satisfies
trusted effect bounds.
The function \lstinline{exchange.swap()} is an \emph{auto-endorsing}~\cite{serif21} function,\footnote
{
  Our definition of an auto-endorsing function is more generous than
  that of \cite{serif21}.  We say a function is
  auto-endorsing either when it updates trusted state itself, or
  when it calls another auto-endorsing function.
}
which can be called in an untrusted context but updates
the trusted state of the exchange.
The function \lstinline{adv.reenter()} is also
an auto-endorsing function because it calls the auto-endorsing
function \lstinline{exchange.swap()}.

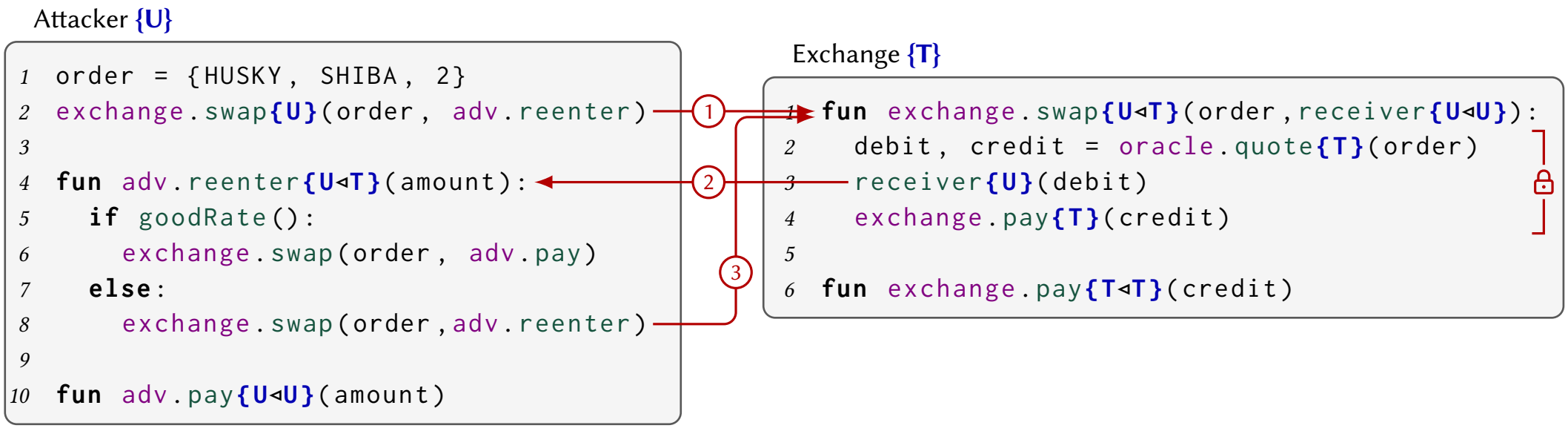
\begin{figure}[t]
    \centering
    \captionsetup{skip=4pt}
    \begingroup
\newcommand{\reMark}[1]{\tikz[remember picture,overlay]\coordinate (#1);}
\lstset{
    basicstyle=\scriptsize\ttfamily,
    numberstyle=\tiny\itshape,
    keywordstyle=[2]\bfseries\color{black},
    keywordstyle=[4]\color[rgb]{0.08,0.32,0.24},
    frame=none,
    aboveskip=0pt,
    belowskip=0pt,
    morekeywords=[3]{exchange, oracle, adv},
    morekeywords=[4]{swap, quote, pay, receiver, goodRate},
    deletedelim=[s]{\{}{\}},
    moredelim=[s][\bfseries\color{darkBlue}]{\{U}{\}},
    moredelim=[s][\bfseries\color{darkBlue}]{\{T}{\}}
}
\scriptsize
\begin{tikzpicture}[
    remember picture,
    >=latex,
    codebox/.style={
        draw=black!65,
        fill=black!4,
        line width=0.5pt,
        rounded corners=3pt,
        inner sep=3pt,
        outer sep=0pt
    },
    boxlabel/.style={
        font=\scriptsize\sffamily,
        fill=none,
        inner xsep=2pt,
        inner ysep=1pt,
        anchor=south west,
        yshift=1pt
    },
    callarrow/.style={->, semithick, rounded corners=3pt,
        draw=red!70!black, shorten <=1pt, shorten >=1pt},
    callnumber/.style={circle, line width=0.6pt, draw=red!70!black, text=red!70!black,
        solid, fill=white, inner sep=0.5pt, minimum size=8pt,
        font=\tiny\sffamily}
]
    \node[codebox, anchor=north west] (re-attacker) at (0, 0) {%
        \begin{minipage}{0.42\linewidth}
            \vspace*{2pt}
            \lstinputlisting{figs/code/Reentrancy-exchange-adv.code}
        \end{minipage}%
    };
    \node[codebox, anchor=north east] (re-exchange) at (\linewidth, {-\baselineskip-1pt}) {%
        \begin{minipage}{0.50\linewidth}
            \vspace*{2pt}
            \lstinputlisting{figs/code/Reentrancy-exchange.code}
        \end{minipage}%
    };

    \node[boxlabel] at ([xshift=5pt]re-attacker.north west)
        {Attacker {\bfseries\color{darkBlue}\{U\}}};
    \node[boxlabel] at ([xshift=5pt]re-exchange.north west)
        {Exchange {\bfseries\color{darkBlue}\{T\}}};

    \coordinate (re-lock-top) at
        ([xshift=-5pt,yshift=6pt]re-exchange.east |- re-lock-start);
    \coordinate (re-lock-bottom) at
        ([xshift=-5pt,yshift=-2pt]re-exchange.east |- re-lock-end);
    \path (re-lock-top) -- coordinate[midway] (re-lock-mid) (re-lock-bottom);
    \coordinate (re-lock-icon) at ([yshift=-1.1pt]re-lock-mid);
    \draw[draw=red!70!black, line width=0.6pt]
        ([xshift=-3pt]re-lock-top) -- (re-lock-top)
        -- ([yshift=5.5pt]re-lock-icon);
    \draw[draw=red!70!black, line width=0.6pt]
        ([yshift=-3.2pt]re-lock-icon)
        -- (re-lock-bottom) -- ([xshift=-3pt]re-lock-bottom);
    \begin{scope}[shift={(re-lock-icon)}, draw=red!70!black,
        fill=red!70!black, line width=0.6pt]
        \draw (-1.4pt,1.5pt) -- (-1.4pt,2.6pt)
            .. controls (-1.4pt,4.5pt) and (1.4pt,4.5pt) .. (1.4pt,2.6pt)
            -- (1.4pt,1.5pt);
        \filldraw[fill=black!4, rounded corners=0.5pt]
            (-2.2pt,-1.7pt) rectangle (2.2pt,1.5pt);
        \fill (0,-0.1pt) circle[radius=0.45pt];
    \end{scope}

    \draw[callarrow]
        ([yshift=2pt]re-initial-call)
        -- node[pos=0.35, callnumber] (re-call-one) {1}
        ([yshift=2pt]re-swap);
    \draw[callarrow]
        ([yshift=2pt]re-receiver)
        -- ([yshift=2pt]re-reenter);
    \node[callnumber]
        at ([yshift=2pt]re-call-one.center |- re-receiver) {2};
    \draw[callarrow]
        ([yshift=2pt]re-recursive-call)
        -- ([xshift=-7pt,yshift=2pt]re-exchange.west |- re-recursive-call)
        -- node[pos=0.25, callnumber] {3}
        ([xshift=-7pt,yshift=0.5pt]re-exchange.west |- re-swap)
        -- ([yshift=0.5pt]re-swap);
\end{tikzpicture}
\endgroup
    \caption{Reentrancy attack against the exchange service.
    The exchange service ``locks'' the trusted control flow level
    on call \#1.
    Call \#2 sends control flow back to untrusted code,
    but the lock prevents call \#3 from reentering trusted control flow.
    }
    \label{fig:reentrancy}
\end{figure}

Auto-endorsement, like any form of endorsement, carries some danger
because it allows an untrusted control-flow context to influence
trusted control flow and state. However, such functions are
indispensable for building software, including smart contracts.
As another example, untrusted code (applications) must be able to
influence trusted (kernel) state via operating system calls.

As observed by Cecchetti et al.~\cite{serif21},
reentrancy attacks are caused by unconstrained use of
auto-endorsing functions.
Bounding label annotations on function pointers enable static
analysis to identify auto-endorsing functions, which prevents these attacks.

Similar to CDA, reentrancy attacks can be prevented by static checks
in the ideal world where attackers are well-typed:
on line 2 of \cref{fig:reentrancy}, the attacker uses
\lstinline{adv.reenter}, a \lstinline|{U$\topbounded$T}|-function,
as an argument of type \lstinline|{U$\topbounded$U}|.
But an ill-typed, real world attacker
may lie about its function signature by publishing \lstinline|{U$\topbounded$U}|
as the signature of \lstinline{adv.reenter}.
The dynamic check that allows the ideal world and the real world to
agree is a dynamic lock mechanism, adapted from Cecchetti et
al.~\cite{serif21}: when a function is called, it locks its own level
of control flow until it returns.
In the example from \cref{fig:reentrancy}, the exchange service locks
control-flow level \lstinline{T} from lines 2 to 4.
Any function call that enters \lstinline{T} from \lstinline{U} is
rejected, so execution aborts the first time \lstinline{adv.reenter()}
tries to invoke \lstinline{exchange.swap()}.

%
%
%
%
%
%

\subsection{Control-Flow Hijacking}

\Cref{fig:base-confusion} extends the same exchange code with
a validation step.
If the validation fails, it executes the error continuation
\lstinline{onReject} that never returns, preventing the 
exchange from happening.

For well-typed attackers, the type of \lstinline{onReject} ensures
that the exchange is only executed when the order validates.
Ill-typed attackers can bypass validation by simply providing
a function \lstinline{adv.ignore} with a unit return type, so
the order keeps executing.

Here, type confusion is used on the ordinary unlabeled part of the type.
Reasoning about such attacks requires specifying how values
are interpreted at incompatible types---for example, what happens
when an integer is used as a function pointer.
We keep the cast semantics abstract and assume that honest code
can inspect the declared ordinary signatures of function pointers
and reject incompatible pointers eagerly at run time.
We leave a systematic study of the broader trade-offs among
cast semantics, runtime assumptions, and dynamic enforcement
to future work.


\begin{figure}[t]
    \centering
    \captionsetup{skip=4pt}
    \begingroup
\newcommand{\rejectMark}[1]{\tikz[remember picture,overlay]\coordinate (#1);}
\lstset{
    basicstyle=\scriptsize\ttfamily,
    numberstyle=\tiny\itshape,
    literate={},
    keywordstyle=[1]\bfseries\color{darkBlue},
    keywordstyle=[2]\bfseries\color{black},
    keywordstyle=[4]\color[rgb]{0.08,0.32,0.24},
    frame=none,
    aboveskip=0pt,
    belowskip=0pt,
    morekeywords=[1]{Nat, Unit, Never},
    morekeywords=[2]{return},
    morekeywords=[3]{exchange, oracle, adv},
    morekeywords=[4]{swap, quote, pay, receiver, validate, onReject, ignore},
    deletedelim=[s]{\{}{\}}
}
\begin{tikzpicture}[
    remember picture,
    >=latex,
    codebox/.style={
        draw=black!65,
        fill=black!4,
        line width=0.5pt,
        rounded corners=3pt,
        inner sep=3pt,
        outer sep=0pt
    },
    boxlabel/.style={
        font=\scriptsize\sffamily,
        fill=none,
        inner xsep=2pt,
        inner ysep=1pt,
        anchor=south west,
        yshift=1pt
    },
    callarrow/.style={->, semithick, rounded corners=3pt,
        draw=red!70!black, shorten <=1pt, shorten >=1pt},
    callnumber/.style={circle, line width=0.6pt, draw=red!70!black, text=red!70!black,
        solid, fill=white, inner sep=0.5pt, minimum size=8pt,
        font=\tiny\sffamily}
]
    \node[codebox, anchor=north west] (reject-attacker) at (0, 0) {%
        \begin{minipage}{0.43\linewidth}
            \vspace*{2pt}
            \lstinputlisting{figs/code/Base-reject-adv.code}
        \end{minipage}%
    };
    \node[codebox, anchor=north east] (reject-exchange) at (\linewidth, 0) {%
        \begin{minipage}{0.49\linewidth}
            \vspace*{2pt}
            \lstinputlisting{figs/code/Base-reject-exchange.code}
        \end{minipage}%
    };

    \node[boxlabel] at ([xshift=5pt]reject-attacker.north west)
        {Untyped attacker {\bfseries\color{darkBlue}\{U\}}};
    \node[boxlabel] at ([xshift=5pt]reject-exchange.north west)
        {Exchange with validation {\bfseries\color{darkBlue}\{T\}}};

    \draw[callarrow]
        ([yshift=2pt]reject-initial-call)
        -- node[midway, callnumber] (reject-call-one) {1}
        ([yshift=2pt]reject-swap);
    \draw[callarrow]
        ([yshift=2pt]reject-on-reject)
        -- ([yshift=2pt]reject-ignore);
    \path let \p1=([yshift=2pt]reject-on-reject), \p2=([yshift=2pt]reject-ignore),
              \p3=(reject-call-one.center) in
        node[callnumber] at ({\x3},
            {\y1 + (\y2-\y1)/ifthenelse(abs(\x2-\x1)<0.01,1,\x2-\x1)*(\x3-\x1)}) {2};
    \draw[callarrow]
        ([yshift=2pt]reject-receiver)
        -- ([yshift=2pt]reject-pay);
    \path let \p1=([yshift=2pt]reject-receiver), \p2=([yshift=2pt]reject-pay),
              \p3=(reject-call-one.center) in
        node[callnumber] at ({\x3},
            {\y1 + (\y2-\y1)/ifthenelse(abs(\x2-\x1)<0.01,1,\x2-\x1)*(\x3-\x1)}) {3};
\end{tikzpicture}
\endgroup
    \caption{Ordinary-type confusion bypasses order validation through a returning
        error continuation. 
        A well-typed argument for \lstinline{onReject} is a function that either
        diverges or aborts, but does not return.
        }
    \label{fig:base-confusion}
\end{figure}

\endgroup

\section{Type-Confusion Security}
\label{sec:formal-theorems}

The examples from \cref{sec:motivating} show that an attacker
who lies about types can cause program behavior that is not
possible for an attacker who obeys the type system.
Our proposed dynamic checks abort at run time to ensure that the real
attacker cannot introduce new behavior.
Type-confusion security requires that all real attackers 
can be _simulated_ by ideal attackers.

In the ideal world,
both the honest party and the attacker are well-typed programs and no
run-time checks are inserted.
Whatever properties the static type system enforces must 
hold in the presence of the attacker.
In the real world, the attacker is an unrestricted
state machine that can produce
arbitrary ill-typed values. To protect against this strong
attacker, the program must in general be translated, to
instrument it with dynamic enforcement mechanisms.

\newcommand{\validR}{\vdash_{\text{R}}}
\newcommand{\validI}{\vdash_{\text{I}}}
\newcommand{\Dyn}[1]{\mathsf{Dyn}_{#1}}
\newcommand{\DynR}{\Dyn{R}}
\newcommand{\DynI}{\Dyn{I}}

Type-confusion security is defined by quantifying over all valid real
attackers, and for each valid real attacker, there must be a valid
ideal attacker. We write $\validR \realattackerconfig$ to mean that
$\realattackerconfig$ is a valid real attacker, and
$\validI \idealattackerconfig$ to mean that
$\idealattackerconfig$ is a valid ideal attacker (for example,
it needs to type-check).

\begin{definition}[Type-Confusion Security]
  \label{def:type-confusion-security-abstract}
  Consider the real world $R = (\validR, \DynR)$,
  and the honest programs' dynamic checks $\DynR$.
  The ideal world is $I = (\validI, \DynI)$,
  with ideal-program dynamic checks $\DynI$.
  Let $\bracketof{\program, \attackerconfig,\Dyn{W}}$
  denote the behavior of the program linked with the attacker $\attackerconfig$
  under $\Dyn{W}$.
  We define \emph{type-confusion security} as follows:
  \[
  R \simulates I
  \iff
    \forall\,\program .\quad
    \forall\,\realattackerconfig |
    \validR \realattackerconfig .\quad
    \exists\,\idealattackerconfig |
    \validI \idealattackerconfig .\quad
    \bracketof{\program, \realattackerconfig,\DynR}
    =
    \bracketof{\program, \idealattackerconfig,\DynI}
  \]
  We write $R \approx I$ when the worlds simulate each other: that is,
  $R\simulates I$ and $I \simulates R$.

\end{definition}

\subsection{Proving Type Confusion by Composing Simulations}
\label{sec:composing-simulations}

\begin{wrapfigure}[14]{r}{0.45\linewidth}
\vskip -1.5em
    \includegraphics[scale=0.43]{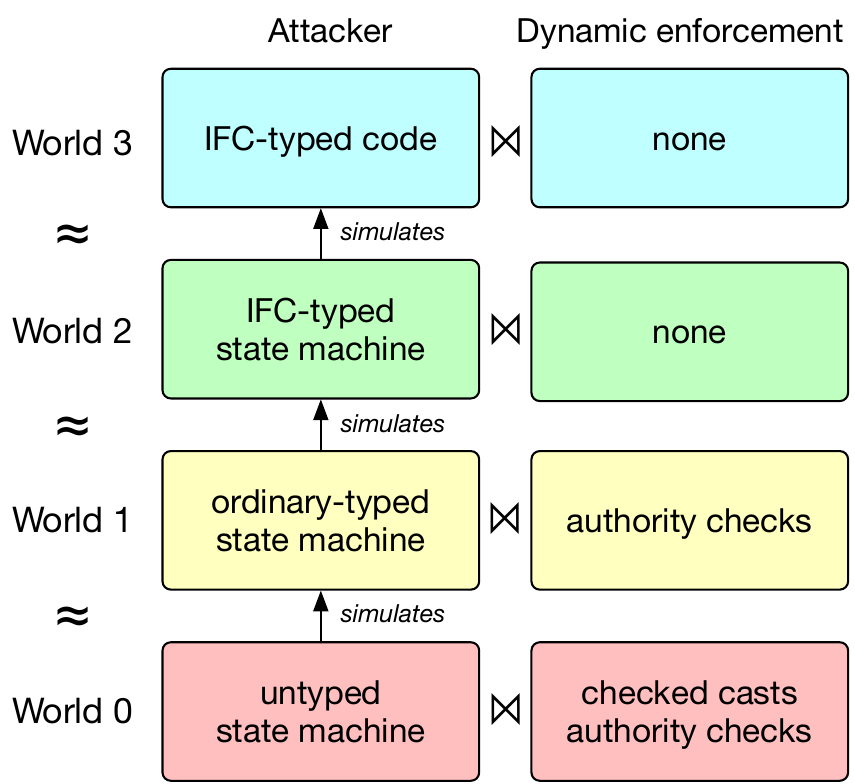}
    \caption{The four worlds and their attackers.}
    \label{fig:worlds-overview}
\end{wrapfigure}

Our proof follows the strategy of context-based
backtranslation~\cite{10.1145/2951913.2951941,Devriese2017}: for every real attacker,
we construct a _simulator_, which is an ideal attacker that simulates the
observable behavior of the real attacker.
As simulation-based security is transitive, multiple
simulators can be composed to construct a bigger simulator.
\Cref{fig:worlds-overview} visualizes our end-to-end type confusion proof, which
is built from 3 smaller type confusion proofs connecting 4 worlds:
Worlds 0--3.

World 3 is the most ``ideal'' world. Both programs and attackers are implemented
as well-typed code from our core calculus, and no dynamic security enforcement
is needed.
To avoid restricting the computational power of attackers to dynamically make decisions based
on run-time values, such as inspecting security types, 
we assume ideal attackers can access a set of _well-typed_ pure functions that may
arbitrarily transform their input values.
In World 0, the most ``real'' world, an attacker is an arbitrary state machine that
does not respect types, and multiple dynamic enforcement mechanisms
are needed.
Worlds 1 and 2 give the attacker powers intermediate
between Worlds 0 and 3.
In World 1, the attacker is a state machine that behaves however it
wants internally but behaves at its interface in a way that respects
ordinary types (but not security labels).
In World 2, the attacker is a state machine that behaves however
it likes internally but fully respects security types at its
interface.

Ultimately, we prove type confusion holds between World 0 and World 3,
with Worlds 1 and 2 as stepping stones.
The most challenging part of the proof is to show that World 1 simulates World 2;
this step turns dynamic IFC enforcement of complex security
properties like reentrancy security and CDA security into purely
static checks.
Simulation between World 2 and World 3 shows that the attacker who
is limited to well-typed code (plus additional well-typed pure functions)
is just as powerful as a state-machine attacker.

By transitivity, the well-typed code
attacker is as powerful as the arbitrary untyped state machine
attacker, assuming that the World 0 dynamic enforcement mechanisms
of safe casts and authority checks are in place at run time.

While our major result is the simulation from World 0 to World 3,
the dynamic checks actually ensure that the backward simulation holds
too:
World 3 simulates World 0.
It might seem trivial that backward simulation holds
because the real attacker is strictly stronger.
However, the backward simulation result is meaningful because an
overly restrictive run-time check that rejects all function calls
might make the main bottom-to-top simulation hold vacuously.
The top-to-bottom, ideal-to-real simulation shows
that the main result is not vacuous.

\subsection{Threat Model}

We consider a decentralized system with principals,
some of which are controlled by the attacker.
Each principal publishes its claimed public interface, which is a set
of function pointers with their signatures.
Implementations of functions on well-typed attackers and honest
principals comply with the published signatures, but
functions implemented by attackers do not follow the signatures.
%

A subtle but important difference between
our threat model and those of prior work in 
gradual typing~\cite{quest-complete, hlio} and dynamic 
IFC~\cite{rajaniAccessControlCapabilities2016,serif21,fabric09}
is the degree of trust in the runtime system.
Prior work designs dynamic mechanisms for a
trusted runtime that also governs the attacker's behavior.
In contrast, we model a system with a minimalist runtime that
cannot control the attacker, and instead ensure that security checks happen 
exclusively in honest code.
Specifically, we assume the following.
First, the callee in a remote call learns who the caller is,
what the arguments are, and nothing else.
Second, all function pointers are publicly callable,
so there is no mechanism for restricting the visibility of function
pointers.
Finally, all static analysis is modularly enforced per function and may
refer to the potentially incorrect function signature table.

\subsection{Information Flow Labels}
\label{sec:ifc-background}

Intuition for why CDA and reentrancy attacks
are the _only_ kinds of IFC type confusion
can be obtained by taking a closer look at IFC labels
and the subtyping lattice of IFC types.

\subsubsection{IFC Label Model}
\label{sec:ifc-label-model}

In an IFC type system, each type $\type$ includes a
label that governs the use of values of that type~\cite{sm-jsac}.
In this paper, we adopt the label model of \citet{ifc-delegation, flam},
where each principal $\p \in \p!$ is also a security policy label $\lbl$.
A value annotated by $\p$ is trusted by the principal $\p$.
Labels form a bounded distributive lattice, where join $\join$ and meet $\meet$
take the greatest common integrity and least combined integrity respectively.
The attacker controls a _prime filter_ \cite{ifc-delegation} of the lattice.\footnote
{
  The attacker controls a non-empty, upward-closed set
  of principals.
  An attacker controls the least combined authority of all principals it controls.
  If two principals are not controlled by
  the attacker, neither is their greatest common authority.
}
Given a type $\type$, we write $\labelof\type$
to denote the label of the type.
We do not consider confidentiality.

The \emph{flows-to} order $\flowsto$ between labels represents the direction of
secure information flow. Ordering $\lbl_1 \flowsto \lbl_2$ holds when $\lbl_1$ is at least
as trusted as $\lbl_2$. The bottom element $\bot$ represents the most
trusted label never controlled by the attacker, and the top element $\top$ represents the 
least trusted label always controlled by the attacker.

\subsubsection{IFC Function Signatures}
\label{sec:ifc-signature}

Function signatures have the form
$\funsig{\type_i}{\type_o}{\pcex}{\pctop}$, where
$\type_i$ is the input type and $\type_o$ is the output type.
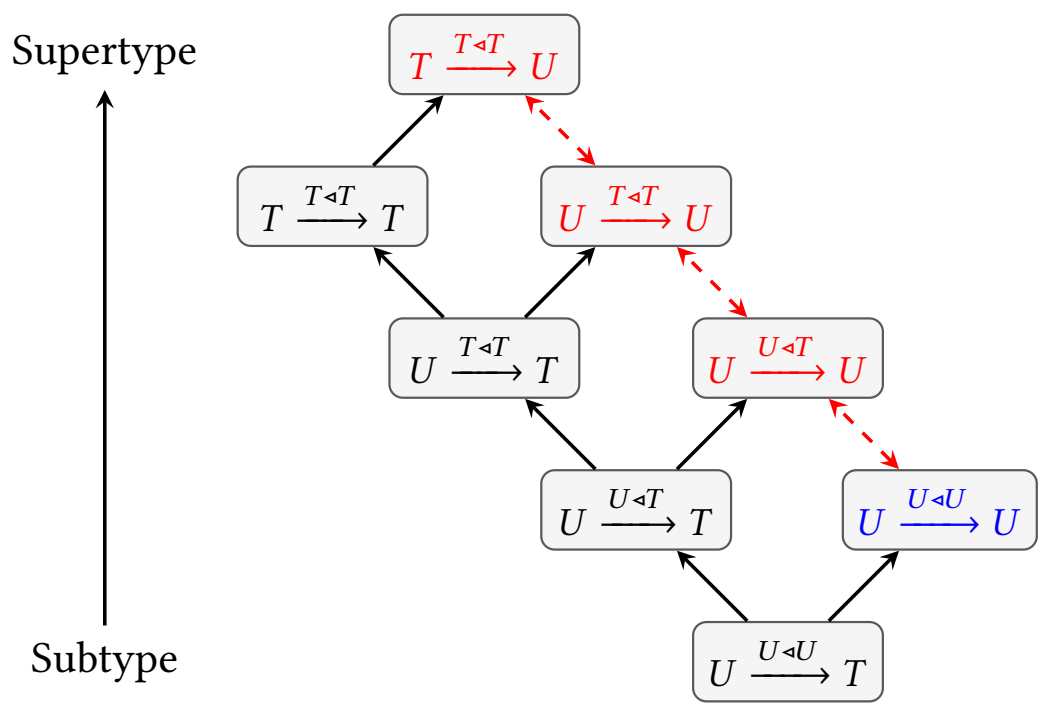
\begin{wrapfigure}{L}{0.5\textwidth}
    \centering
    \resizebox{\linewidth}{!}{\begin{tikzpicture}[
  node/.style={draw=black!65, fill=black!4, rounded corners=3pt,
    line width=0.5pt, align=center, font=\small,
    minimum width=1.5cm, minimum height=0.6cm},
  arrow/.style={->, thick, >=stealth},
  scale=0.6
]

\node[node, text=red] (Z) at (-2,5) {$T \xrightarrow{T \topbounded T} U$};
\node[node, text=red] (A) at (0,3) {$U \xrightarrow{T \topbounded T} U$};
\node[node, text=red] (B) at (2,1) {$U \xrightarrow{U \topbounded T} U$};
\node[node, text=blue] (C) at (4,-1) {$U \xrightarrow{U \topbounded U} U$};

\node[node] (F) at (-4,3) {$T \xrightarrow{T \topbounded T} T$};
\node[node] (G) at (-2,1) {$U \xrightarrow{T \topbounded T} T$};
\node[node] (H) at (0,-1) {$U \xrightarrow{U \topbounded T} T$};
\node[node] (I) at (2,-3) {$U \xrightarrow{U \topbounded U} T$};

\draw[arrow] (I) -- (H);
\draw[arrow] (H) -- (G);
\draw[arrow] (G) -- (F);

\draw[arrow] (F) -- (Z);
\draw[arrow] (G) -- (A);
\draw[arrow] (H) -- (B);
\draw[arrow] (I) -- (C);

\draw[arrow, <->, dashed, red] (A) -- (Z);
\draw[arrow, <->, dashed, red] (B) -- (A);
\draw[arrow, <->, dashed, red] (C) -- (B);

\node (sup) at (-7,5) {Supertype};
\node (sub) at (-7,-3) {Subtype};
\draw[arrow] (sub) -- (sup);

\end{tikzpicture}}
    \caption{The subtyping lattice of well-formed function types.
    Type confusion occurs when a down-cast happens to a fully
    untrusted pointer.
    }
  \label{fig:subtype-lattice-all}
\end{wrapfigure}
As introduced in \cref{sec:motivating}, the external pc $\pcex$ 
upper-bounds the contexts in which the function may be called;
the bounding pc $\pctop$ upper-bounds the effect of the function
starting from its invocation to its return.
Like other types, function types $\type =
\funptrtype{\type_i}{\type_o}{\pcex}{\pctop}{\lbl}$
include an extra label $\lbl$, the label of the function pointer itself.

In a decentralized system, an IFC-labeled function type 
is a security specification.
As certain security specifications can be self-contradictory,
such function types are considered ill-formed.
Let $\p$ be the principal on which this function is defined.
A well-formed function type is one where
$\pctop \flowsto \p \flowsto \pcex \flowsto \labelof{\type_i}$.
First, because each function runs at its own control-flow integrity,
the upper bound on effects $\pctop$ needs to be at least as trusted as $\p$ itself.
Second, a principal should be allowed to call its own function,
so $\p$ needs to be no less trusted than $\pcex$.
Finally, $\pcex \flowsto \labelof{\type_i}$ should hold;
if not, this function would take trusted input from an untrusted context
that taints any value it holds, causing a contradiction.

\Cref{fig:subtype-lattice-all} shows the subtyping lattice of 
IFC function types. As the lattice shows,
the output label $\type_o$ is covariant;
the input type $\type_i$, external pc $\pcex$, 
and bounding pc $\pctop$ are all contravariant.

\subsubsection{Type Confusion of IFC Function Types}
\label{sec:ifc-type-confusion}

In trusted code, the only way to statically type a _useful_ function pointer
received from an attacker is the type
\smash{$\funptrtype{U}{U}{U}{U}{U}$}.
To see why, first observe that all pointers
from the attackers are $U$-labeled.
This label taints all labels except $\pctop$ in
a function pointer \cite{zm01, serif21}.
%
Therefore, the type system gives all
attacker-controlled pointers the type
\smash{$\funptrtype{U}{U}{U}{U}{U}$}
or \smash{$\funptrtype{U}{U}{U}{T}{U}$}.

In a trusted context, pointers of the second type can be passed or returned,
but cannot be directly invoked because invoking them leads to a 
potential reentrancy attack.
Thus only the first kind can be used in a trusted context.

Therefore, IFC-type confusion only happens when an attacker performs a 
down-cast to a \smash{$\funptrtype{U}{U}{U}{U}{U}$}-typed pointer, and then
passes the pointer to the honest principal (deputy).
As shown in \cref{fig:subtype-lattice-all}, there are
three possible down-casts, one for each contravariant label.
CDAs and reentrancy attacks are the cases where down-casts happen
on $\pcex$ and $\pctop$.
Type confusion on primitive (non-function) input types can be statically
enforced because the deputy treats inputs of such pointers
as untrusted.
In the case where the input type is itself another function type,
well-formedness of IFC types 
statically ensures that no such type confusion is possible.

\section{A Core Calculus for Type Confusion}

\label{sec:calculus}

We define a core calculus to formally
model a decentralized system that supports first-class
function pointers and remote function calls.
We represent honest principals by their program code and the attacker by an
interactive state machine, which is linked to the program at run time.

\subsection{Syntax for the Honest Party}
\label{sec:syntax}

\begin{figure}
  \centering
  \[
  \begin{array}{@{}rcl@{\qquad}l@{}}
    p,q,\ell,\pc &\in& \bbP
      & \text{principals and labels}\\
    f &\in& \FunName
      & \text{function names}\\
    h &\in& \RefName
      & \text{cell names}\\

    \tau &::=&
      \tboolof{\ell}
      \mid \tnatof{\ell}
      \mid \tunit
      \mid \tempty
      \mid \funptrtype{\tau_i}{\tau_o}{\pcex}{\pctop}{\ell}
      & \text{types}\\
    \Signature &\ni&
      \funsig{\tau_i}{\tau_o}{\pcex}{\pctop}
      & \text{signatures}\\
    v &::=&
      \unit \mid b \mid n \mid \funid{p}{f} \mid x
      & \text{values}\\
    e &::=&
      v \mid \abort \mid \exprop{v_1}{v_2}
      \mid \pcapp{v_1}{v_2}{\ell}
      \mid \exprdowngrade{v}{\ell}
      & \text{expressions}\\
      &&
      \mid \writeexpr{h}{v} \mid \readexpr{h}
      \mid \exprlet{x}{e_1}{e_2}
      \mid \exprif{v}{e_1}{e_2}
      &\\
    \Body &\ni& \lam{x}{e}
      \quad\text{where } \mathit{fv}(e)\subseteq\{x\}
      & \text{function bodies}\\

    \event &::=&
      \writeevent{p}{h}{v}
      \mid \eventabort
      \mid \eventdone
      \mid \eventcall{\funid{p}{f}}{\ell}
      \mid \eventret{p}
      & \text{events; traces } \trace
  \end{array}
  \]
  \caption{Syntax of the core calculus.  Principals and labels range over
  the same lattice $\bbP$.}
  \label{fig:core-syntax}
\end{figure}

\Cref{fig:core-syntax} defines the syntax of the core calculus.  This language models a
decentralized system with multiple principals $p\in\bbP$. 
Each principal $p$ can also be interpreted as a label $\ell$,
which can be used to describe the integrity of values.
Principals own functions and memory cells.  We write $\funid{p}{f}$ for
the identifier, or pointer, of the function $f$ owned by principal $p$.

A program $\calP=(\coreiface,\bodytable,\funid{\top}{\entry})$ consists
of an interface, a body table and an entry function belonging to
principal $\top$.
The interface $\coreiface=(\sigtable,\celltable)$ 
pairs a signature table $\sigtable:\bbP \times
\FunName\finpartial\Signature$, which maps each
declared function to a signature, with a cell table
$\celltable:\bbP\times\RefName\finpartial\Type$, which maps each
cell to the type of its contents.
%
%
The body table $\bodytable:\bbP\times\FunName\finpartial\Body$ maps each implemented
function to a closed lambda term (i.e., $\lambda\;x.e$), where $\cdot\finpartial \cdot$ denotes a finite partial map.
We use $\funid{p}{f}$ and $\funid{p}{h}$ to range over elements of
$\bbP\times\FunName$ and $\bbP\times\RefName$ respectively.
Every execution starts at $\funid{\top}{\entry}$, which belongs to the
least trusted principal $\top$ with signature 
\smash{$\funsig{\tunit}{\tunit}{\top}{\bot}$}.
Thus, all executions are initiated by the attacker.

Programs must satisfy two well-formedness conditions. First, all three tables have finite
domains, because a program is a finite artifact. Second,
$\dom(\bodytable)\subseteq\dom(\sigtable)$, so every implemented function has a
published signature. Note that the converse need not hold, an honest party does
not know the implementation of an attacker in general (\cref{sec:attackers}).

%
%
%

%

Types are equipped with information-flow labels in the standard
way~\cite{sm-jsac,zm03}.
Booleans, natural numbers, and function pointers each carry a label
$\ell$ denoting the integrity policy of the label.
The unit type $\tunit$ and the empty type $\tempty$ have at most one
inhabitant, and thus do not require labeling.

Values and expressions are mostly standard and written in A-normal
form~\cite{flanagan93}.
The language aims to be practical, and thus supports
endorsement~\cite{iflow-properties} of non-function-pointer values
$v\mathbin{\downarrow}\ell$ as strict noninterference is too restrictive.
Endorsement of function pointers is disallowed because it does
not add meaningful expressiveness, and it adds a new surface for type-confusion attacks.
A principal can only read directly from and write directly to a cell $h$ that it owns;
remote read and write must be provided by ``getter'' and ``setter'' functions.
A function application $\pcapp{v_1}{v_2}{\ell}$ is annotated with the pc
$\ell$ used for this function call.
In well-typed code, $\ell$ can be inferred and is equal to the
external pc of pointer $v_1$'s function type. 
We add it to the syntax explicitly because it is used by run-time checks.

As a program executes, it produces a trace $\trace$ of events $\event$.
We write $::$ for list concatenation, treating individual elements as singleton lists.
A write event $\writeevent{p}{h}{v}$ records that $p$ writes $v$ to its
cell $h$, and a call event $\eventcall{\funid{p}{f}}{\ell}$ records a call to $\funid{p}{f}$
together with its calling pc $\ell$, the caller's label joined with the
call's annotation.  A return event $\eventret{p}$ records that a function
owned by $p$ returns.  Finally, $\eventabort$ records that the run aborts,
and $\eventdone$ records that it exits gracefully once the entry returns.

\subsection{Static Semantics}
\label{sec:static-semantics}

\begin{figure*}
  \centering
  \small
  \begin{judgments}
    \judgment{\tau\subtypeof\tau'}
  \end{judgments}
  \begin{mathpar}[\mprset{lineskip=.8em,andskip=1.25em,sep=1em}]
    \inferrule*[Lab=S-Unit]
      { }
      {\tunit\subtypeof\tunit}
    \and
    \inferrule*[Lab=S-Empty]
      { }
      {\tempty\subtypeof\tau}
    \and
    \inferrule*[Lab=S-Bool]
      {\ell\flowsto\ell'}
      {\tboolof{\ell}\subtypeof\tboolof{\ell'}}
    \and
    \inferrule*[Lab=S-Nat]
      {\ell\flowsto\ell'}
      {\tnatof{\ell}\subtypeof\tnatof{\ell'}}
    \and
    \inferrule*[Lab=S-Arrow]
      {\ell\flowsto\ell'
       \and \pcex'\flowsto\pcex
       \and \pctop'\flowsto\pctop\\
       \tau'_i\subtypeof\tau_i
       \and \tau_o\subtypeof\tau'_o}
      {\funptrtype{\tau_i}{\tau_o}{\pcex}{\pctop}{\ell}
       \subtypeof
       \funptrtype{\tau'_i}{\tau'_o}{\pcex'}{\pctop'}{\ell'}}
  \end{mathpar}

  \begin{judgments}
    \judgment{\ctyping{\Gamma}{\pc}{p}{\pcself{\text{max}}}{e}{\tau}}
  \end{judgments}
  \begin{mathpar}
    \inferrule*[Lab=TI-Fun]
      {\sigtable(\funid{q}{f})=
         \funsig{\tau_i}{\tau_o}{\pcex}{\pctop}
       \and q\flowsto\ell}
      {\ctyping{\Gamma}{\pc}{p}{\pcself{\text{max}}}
        {\funid{q}{f}}
        {\funptrtype{\tau_i}{\tau_o}{\pcex}{\pctop}{\ell}}}
    \and
    \inferrule*[Lab=TI-Downgrade]
    {\exprdowngrade{\tau}{\ell}\text{ is defined}\\\\
      \ctyping{\Gamma}{\pc}{p}{\pcself{\text{max}}}{v}{\tau}
       }
      {\ctyping{\Gamma}{\pc}{p}{\pcself{\text{max}}}
        {\exprdowngrade{v}{\ell}}{\exprdowngrade{\tau}{\ell}}}
    \and
    \inferrule*[Lab=TI-Read]
      {\celltable(p,h)=\tau
       \and \pc\flowsto\labelof{\tau}}
      {\ctyping{\Gamma}{\pc}{p}{\pcself{\text{max}}}{\readexpr{h}}{\tau}}
    \and
    \inferrule*[Lab=TI-Write]
      {\celltable(p,h)=\tau \and \pc\flowsto\labelof{\tau}\\\\
       \ctyping{\Gamma}{\pc}{p}{\pcself{\text{max}}}{v}{\tau'}
       \and\tau'\subtypeof\tau
       }
      {\ctyping{\Gamma}{\pc}{p}{\pcself{\text{max}}}
        {\writeexpr{h}{v}}{\tunit}}
    \and
    \inferrule*[Lab=TI-App]
      {\ctyping{\Gamma}{\pc}{p}{\pcself{\text{max}}}{v_1}
         {\funptrtype{\tau_i}{\tau_o}{\pcex}{\pctop}{\ell}}\\\\
       \ctyping{\Gamma}{\pc}{p}{\pcself{\text{max}}}{v_2}{\tau}
       \and \tau\subtypeof\tau_i\\\\
       \pc\join\ell\flowsto\pcex
       \and \pcex\flowsto\pctop\join p
       \and \pcself{\text{max}}\flowsto\pctop}
      {\ctyping{\Gamma}{\pc}{p}{\pcself{\text{max}}}
        {\pcapp{v_1}{v_2}{\pcex}}{\tau_o\join\ell}}
  \end{mathpar}
  \caption{Full IFC subtyping and selected expression typing rules.}
  \label{fig:core-typing}
\end{figure*}

\paragraph{IFC typing.}
\Cref{fig:core-typing} defines the judgments
$\ctyping{\Gamma}{\pc}{p}{\pcself{\text{max}}}{e}{\tau}$ and
$\tau\subtypeof\tau'$. The former checks
an expression $e$, run by principal $p$ at pc $\pc$ with bounding pc
$\pcself{\text{max}}$, against type $\tau$.\footnote{The typing
  rules are decidable whenever the label order is decidable, as in
  practical IFC label models~\cite{myers-popl99, viaduct-pldi21, ifc-delegation}.}
The latter gives IFC subtyping rules which are standard.

\Cref{fig:core-typing} shows selected rules, and
\cref{sec:ifc-static-semantics} gives the full type system.
Most rules are standard~\cite{zm01,zm01b}.
Well-typed code can access only addresses declared by $\celltable$.
\ruleof{TI-Read} and \ruleof{TI-Write} access only the declared cells of
$p$, and $\pc\flowsto\labelof{\tau}$ ensures that the cell's label
covers the control flow that reaches the access.
\ruleof{TI-Downgrade} applies to values that are not function-typed
and sets its label to $\lbl$.
In particular, function pointers cannot be endorsed, so their types retain all
taints to them.

Function pointers require more care. Since the owner $q$ controls how
$\funid{q}{f}$ behaves, a pointer to it is trusted only as much as $q$
is, so \ruleof{TI-Fun} requires $\q \flowsto \lbl$.
Combined with the well-formedness conditions of signatures
from \cref{sec:ifc-signature}, 
well-formedness of types, written $\wfcore{\tau}$, is given by:
\[
  \wfcore{\funptrtype{\tau_i}{\tau_o}{\pcex}{\pctop}{\ell}}
  \quad\text{iff}\quad
  \wfcore{\tau_i}\land\wfcore{\tau_o}\land\pctop\flowsto\ell
\]

\ruleof{TI-App} enforces the static discipline of
function calls through three
constraints.  First, $\pc\join\ell\flowsto\pcex$ allows a call only
when the external pc covers both the caller's pc and the influence over which
function the pointer selects.
Second, $\pcex\flowsto\pctop\join p$ ensures that a call enters at an
untrusted level only to the extent that the callee stays that untrusted
throughout or the caller already was, which forbids the control-flow
endorsement behind reentrancy~\cite{serif21}.
Third, the caller's bounding pc constrains every function it calls, so
the constraint $\pcself{\text{max}}\flowsto\pctop$ accepts only callees within such a bound.  
The result type
$\tau_o\join\ell$ joins the label of $\tau_o$ with $\ell$ 
(see \cref{sec:notation-overload} for formal definitions),
since the resulting value reveals which function the
pointer selected.

\paragraph{Ordinary typing.}
Ordinary typing follows the standard type system of the simply-typed
lambda calculus, and is isomorphic to IFC typing over the one-point
label model $\{\bullet\}$.  For brevity, we write
$\sigma::=\realbooltype\mid\realnattype\mid\tunit\mid\tempty
\mid\realfunctiontype{\sigma}{\sigma}$, and
$\rtyping{\Gamma}{p}{e}{\sigma}$ for ordinary typing rules.
We write $\erase{\tau}$ for the unlabeled type that results when every label and pc annotation is erased from $\tau$.
With the one-point label, every label premise in $\vdash_{\mathrm W_2}$ trivially holds, so
only the ordinary-type premises remain. We show \ruleof{TR-Fun},
\ruleof{TR-App} and \ruleof{TR-Down}, and the rest of the type system
can be found in \cref{fig:core-real-typing}.
\[
  \inferrule*[Lab=TR-Fun]
    {\sigtable(\funid{q}{f})=
       \funsig{\tau_i}{\tau_o}{\pcex}{\pctop}}
    {\rtyping{\Gamma}{p}{\funid{q}{f}}
      {\realfunctiontype{\erase{\tau_i}}{\erase{\tau_o}}}}
  \qquad
  \inferrule*[Lab=TR-App]
    {\rtyping{\Gamma}{p}{v_1}
       {\realfunctiontype{\sigma_i}{\sigma_o}}\\\\
     \rtyping{\Gamma}{p}{v_2}{\sigma}
     \and \sigma\subtypeof\sigma_i}
    {\rtyping{\Gamma}{p}{\pcapp{v_1}{v_2}{\pcex}}{\sigma_o}}
  \qquad
  \inferrule*[Lab=TR-Down]
  {\sigma\in\{\realbooltype,\realnattype,\tempty\}\\\\
    \rtyping{\Gamma}{p}{v}{\sigma}
     }
    {\rtyping{\Gamma}{p}{\exprdowngrade{v}{\ell}}{\sigma}}
\]
Note that an IFC-typed expression is also
ordinarily typed 
($\ctyping{\Gamma}{\pc}{p}{\pcself{\text{max}}}{e}{\tau} \implies$
$\rtyping{\erase{\Gamma}}{p}{e}{\erase{\tau}}$).

\paragraph{Program typing.}
An interface is well formed when every cell type and every signature is
well formed (\cref{sec:ifc-signature}).
\[
\begin{aligned}
  \vdash_{\mathrm W_2}\coreiface
  \quad\text{iff}\quad
  &\forall p,f.\;
    \sigtable(\funid{p}{f})=\funsig{\tau_i}{\tau_o}{\pcex}{\pctop}
    \Longrightarrow
    \wfcore{\tau_i}\land\wfcore{\tau_o}\land
    \pctop\flowsto p\flowsto\pcex\flowsto\labelof{\tau_i}\\
  \land\;&\forall p,h.\;
    \celltable(p,h)=\tau\Longrightarrow\wfcore{\tau}
\end{aligned}
\]
A program is IFC-typed when its interface is well-formed, and every
function body is typed against the corresponding signature.
\[
\begin{aligned}
  \vdash_{\mathrm W_2}\calP
  \quad\text{iff}\quad
  \vdash_{\mathrm W_2}\coreiface\land
  \forall p,f.\;
  \left.
  \begin{array}{c}
    \sigtable(\funid{p}{f})=\funsig{\tau_i}{\tau_o}{\pcex}{\pctop}\\
    \bodytable(\funid{p}{f})=\lam{x}{e}
  \end{array}
  \right\}
  \Longrightarrow
  \ctyping{[\tau_i]}{p}{p}{\pctop}{e}{\tau_o\join p}
\end{aligned}
\]
The body runs at pc $p$ with result $\tau_o\join p$, which models
auto-endorsement on entry and is exactly the trust that
$\pcex\flowsto\pctop\join p$ in \ruleof{TI-App} guards.  Interfaces in
World 1 are trivially well-formed, so
$\vdash_{\mathrm W_1}\calP$ holds when every body satisfies the ordinary
typing judgment
$\rtyping{[\erase{\tau_i}]}{p}{e}{\erase{\tau_o}}$.
From above, every
IFC-typed program is also ordinarily typed ($\vdash_{\mathrm W_2} \calP
\implies \vdash_{\mathrm W_1}\calP$).

\subsection{Attackers}
\label{sec:attackers}

\begin{figure}
  \centering
  $\begin{gathered}
    \begin{array}{@{}rcl@{\qquad}l@{}}
      a\in\Action &::=&
        \actioncall{v_f}{\ell}{v} \mid \actionret{v} \mid \actionabort
        & \text{actions}\\
      m\in\Message &::=&
        \messageinvoke{\funid{p}{f}}{v} \mid \messageresult{v}
        & \text{messages}
    \end{array}\\
    \mathcal M=(\State,s_{\calA}^{\mathrm{init}},\rho,\delta)
    \quad\text{where}\quad
    s_{\calA}^{\mathrm{init}} \in \State\\
    \rho:\State\times\Message\to\State
    \quad
    \delta:\State\to\State+(\Action\times\State)
  \end{gathered}$
  \caption{An attacker $\calA=(\adv,\sigtable_{\calA},\mathcal M)$ controls the principals $\adv$, declares
  the signatures $\sigtable_{\calA}$, and behaves as the state machine
  $\mathcal M$, which receives messages via $\rho$ and emits actions
  via $\delta$.}
  \label{fig:core-attacker}
\end{figure}

\paragraph{State Machine Attacker}
\Cref{fig:core-attacker} presents the definition of a state machine attacker
$\calA=(\adv,\sigtable_{\calA},\mathcal M)$, which controls a set of
principals $\adv\subseteq\bbP$ that forms a prime filter,
as explained in \cref{sec:ifc-label-model}.
The attacker's
behavior is given by an interactive state machine $\mathcal M$ with
private state $\State$~\cite{patrignani22uc}.
Whenever control passes to the attacker, the receive function $\rho$ updates the state with
the incoming message, which either invokes an attacker-owned function
$\funid{p}{f}$ on an argument $v$ or returns a result $v$ from an
earlier call. The transition function $\delta$ then either takes an
internal step or emits an action, calling a function $v_f$ at label
$\ell$, returning a value $v$, or aborting. Because $\delta$ is total,
an attacker can spin by only taking internal steps. 
The state machine can model attacker code written in a different language.

Linking an attacker $\calA$ with a program $\calP$ is denoted by $\calP\bowtie\calA$. The
attacker contributes a finite signature table $\sigtable_{\calA}$,
which declares functions owned by principals in $\adv$. The linked
interface $\sigtable\bowtie\sigtable_{\calA}$ extends $\sigtable$ 
but never overrides the program's own,
\[
  (\sigtable\bowtie\sigtable_{\calA})(\funid{p}{f})=
  \begin{cases}
    \sigtable(\funid{p}{f}) & \text{if }\funid{p}{f}\in\dom(\sigtable)\\
    \sigtable_{\calA}(\funid{p}{f}) & \text{otherwise}
  \end{cases}
\]
Thus besides declaring new functions, the attacker can also
replace an existing function, without changing its signature,
if that function belongs to an attacker-controlled principal.
Signatures of attacker
functions are merely claims, because $\mathcal M$ is free to
break them.

\paragraph{Attacker Boundary}
The attacker and the program interact only across the \emph{boundary}
between them, where control enters $\mathcal M$ as a message and leaves
it as an action.  Each world admits a different class of attackers, and
the classes differ only in the typing that $\mathcal M$ must respect at
this boundary.
The rest of this subsection defines the IFC-typed boundary for World 2
and the ordinarily typed boundary for World 1.
Attackers in World 0 are unconstrained and
attackers in World 3 are programs but not state machines.
They are described in
\cref{sec:base-real,sec:ideal-code}, respectively.

A message $m$ or action $a$ is typed in a world $\mathrm W\in\{\mathrm
W_1,\mathrm W_2\}$
against the signature of the function $\funid{p}{f}$ it concerns in the linked
interface $\sigtable\bowtie\sigtable_{\calA}$, written $(\sigtable\bowtie\sigtable_{\calA})(\funid{p}{f})
=\funsig{\tau_i}{\tau_o}{\pcex}{\pctop}$.
\[
\begin{array}{@{}l|l@{\quad}l@{}}
  & \mathrm W=\mathrm W_2 & \mathrm W=\mathrm W_1\\
  \hline
  \vdash_{\mathrm W}\messageinvoke{\funid{p}{f}}{v}
    & p\in\adv\land\ctyping{\emptyset}{p}{p}{\pctop}{v}{\tau_i}
    & p\in\adv\land\rtyping{\emptyset}{p}{v}{\erase{\tau_i}}\\
  \funid{p}{f}\vdash_{\mathrm W}\messageresult{v}
    & \ctyping{\emptyset}{p}{p}{\pctop}{v}{\tau_o\join p}
    & \rtyping{\emptyset}{p}{v}{\erase{\tau_o}}
\end{array}
\]
An action emitted by $\funid{p}{f}$ is well typed when
\[
\begin{array}{@{}l|l@{\quad}l@{}}
  & \mathrm W=\mathrm W_2 & \mathrm W=\mathrm W_1\\
  \hline
  \funid{p}{f}\vdash_{\mathrm W}\actioncall{v_f}{\ell}{v}
    & \exists\pc,\tau.\;p\flowsto\pc\land
      \ctyping{\emptyset}{\pc}{p}{\pctop}{\pcapp{v_f}{v}{\ell}}{\tau}
    & \exists\sigma.\;\rtyping{\emptyset}{p}{\pcapp{v_f}{v}{\ell}}{\sigma}\\
  \funid{p}{f}\vdash_{\mathrm W}\actionret{v}
    & \ctyping{\emptyset}{p}{p}{\pctop}{v}{\tau_o\join p}
    & \rtyping{\emptyset}{p}{v}{\erase{\tau_o}}\\
  \funid{p}{f}\vdash_{\mathrm W}\actionabort
    & \text{always} & \text{always}
\end{array}
\]
A call is thus one that the attacker function could have made from some
program point of a well-typed body, and a return delivers what a
well-typed body of $\funid{p}{f}$ would produce.  The World 1
judgments $\vdash_{\mathrm W_1}$ use ordinary typing against
$\erase{\tau_i}$ and $\erase{\tau_o}$.

\begin{figure}
  \centering
  \small
  \begin{judgments}
    \judgment{\calA\vdash_{\mathrm W}(s,r,K)}
  \end{judgments}
  \begin{mathpar}
    \inferrule*[Lab=R-Init]
      { }
      {\calA\vdash_{\mathrm W}(s_{\calA}^{\mathrm{init}},\text{--},\emptyset)}
    \and
    \inferrule*[Lab=R-Ret]
    { \delta(s)=(\actionret{v},s')\\\\
      \calA\vdash_{\mathrm W}(s,\funid{p}{f},K)}
      {\calA\vdash_{\mathrm W}(s',\text{--},K)}
    \and
    \inferrule*[Lab=R-Call]
    {\delta(s)=(\actioncall{\funid{q}{g}}{\ell}{v},s')\\\\
      \calA\vdash_{\mathrm W}(s,\funid{p}{f},K)
       }
      {\calA\vdash_{\mathrm W}(s',\text{--},(\funid{p}{f},\funid{q}{g})::K)}
    \\
    \inferrule*[Lab=R-Step]
      { \delta(s)=s'\\\\
        \calA\vdash_{\mathrm W}(s,\funid{p}{f},K)}
      {\calA\vdash_{\mathrm W}(s',\funid{p}{f},K)}
    \and
    \inferrule*[Lab=R-Invoke]
    {\vdash_{\mathrm W}\messageinvoke{\funid{p}{f}}{v}\\\\
      \calA\vdash_{\mathrm W}(s,\text{--},K)}
      {\calA\vdash_{\mathrm W}(\rho(s,\messageinvoke{\funid{p}{f}}{v}),\funid{p}{f},K)}
    \and
    \inferrule*[Lab=R-Result]
    {\funid{q}{g}\vdash_{\mathrm W}\messageresult{v}\\\\
      \calA\vdash_{\mathrm W}(s,\text{--},(\funid{p}{f},\funid{q}{g})::K)
       }
      {\calA\vdash_{\mathrm W}(\rho(s,\messageresult{v}),\funid{p}{f},K)}
  \end{mathpar}
  \caption{Reachable configurations of an attacker under typed messages.}
  \label{fig:boundary-reach}
\end{figure}

In \Cref{fig:boundary-reach}, a configuration $(s,r,K)$ pairs the state
$s$ of $\mathcal M$ with the running attacker function $r$, or
$r=\text{--}$ when control is outside, and with a stack $K$ of pending
calls.  A pair $(\funid{p}{f},\funid{q}{g})$ in $K$ records that
$\funid{p}{f}$ called $\funid{q}{g}$ and awaits its result.
\ruleof{R-Invoke} and \ruleof{R-Result} are the only rules that receive
messages, and both require those messages to be well-typed with respect to $\mathrm
W$.
\ruleof{R-Result} delivers a result only to the innermost
pending call, which makes the run well-bracketed in the sense of game semantics~\cite{abramsky99game}.  No rule continues
after an abort.
\begin{definition}[Boundary typing]
  \label{def:boundary-typing}
  An attacker $\calA$ is boundary-typed in $\mathrm W$, written
  $\vdash_{\mathrm W} \calA$, when
  $\calA\vdash_{\mathrm W}(s,\funid{p}{f},K)$ and $\delta(s)=(a,s')$ imply
  $\funid{p}{f}\vdash_{\mathrm W}a$.
\end{definition}
A boundary-typed attacker is never the first to break typing~\cite{abadi95conjoining}.  The
condition constrains $\mathcal A$ only after messages that a
well-typed program can send.

\subsection{Dynamic Semantics}
\label{sec:dynamic-semantics}

\Cref{fig:core-honest-semantics} gives the semantics of honest code in
the standard style of evaluation contexts~\cite{wf94}.  This semantics
leaves every interaction with other principals to the linked semantics,
which ensures that honest code runs the same way in every world (except for dynamic checks).

\begin{figure}
  \centering
  \[
  \begin{array}{@{}rcl@{\qquad}l@{}}
    E &::=& [\cdot]\mid\exprlet{x}{E}{e}
      & \text{evaluation contexts}\\
    N &::=& \internaloutcome{e}\mid\emitoutcome{a}{E}
      & \text{outcomes}
  \end{array}
  \]
  \begin{judgments}
    \judgment{
      \langle e,\store\rangle
      \estep{p}
      \langle N,\store',\trace\rangle}
  \end{judgments}
  \begin{mathpar}[\mprset{lineskip=.8em,andskip=1.2em,sep=1em}]
    \inferrule*[Lab=R-Op]
    {
    }
      {\langle\exprop{v_1}{v_2},\store\rangle
       \estep{p}
       \langle\internaloutcome{v_1 \sem{\otimes} v_2},\store,\epsilon\rangle}
    \and
    \inferrule*[Lab=R-If]
    {
    }
    {\langle\exprif{b}{e_{\true}}{e_{\false}},\store\rangle
      \estep{p}
      \langle\internaloutcome{e_b},\store,\epsilon\rangle}
    \and
    \inferrule*[Lab=R-Down]
      { }
      {\langle\exprdowngrade{v}{\ell},\store\rangle
       \estep{p}
       \langle\internaloutcome{v},\store,\epsilon\rangle}
    \and
    \inferrule*[Lab=R-Let]
      { }
      {\langle\exprlet{x}{v}{e},\store\rangle
       \estep{p}
       \langle\internaloutcome{\subst{e}{x}{v}},\store,\epsilon\rangle}
    \and
    \inferrule*[Lab=R-Read]
      {\store(p,h)=v}
      {\langle\readexpr{h},\store\rangle
       \estep{p}
       \langle\internaloutcome{v},\store,\epsilon\rangle}
    \and
    \inferrule*[Lab=R-Write]
      {\store'=\update{\store}{(p,h)}{v}}
      {\langle\writeexpr{h}{v},\store\rangle
       \estep{p}
       \langle\internaloutcome{\unit},\store',\writeevent{p}{h}{\opaque{v}}\rangle}
    \and
    \inferrule*[Lab=R-Call]
      { }
      {\langle\pcapp{v_f}{v}{\ell},\store\rangle
       \estep{p}
       \langle\emitoutcome{\actioncall{v_f}{\ell}{v}}{[\cdot]},\store,\epsilon\rangle}
  \end{mathpar}
  \begin{judgments}
    \judgment{
      \langle e,\store\rangle
      \xrightarrow{p}
      \langle N,\store',\trace\rangle}
  \end{judgments}
  \begin{mathpar}[\mprset{lineskip=.8em,andskip=1.2em,sep=1em}]
    \inferrule*[Lab=S-Internal]
      {\langle e,\store\rangle
         \estep{p}
         \langle\internaloutcome{e'},\store',\trace\rangle}
      {\langle\evalctxof{e},\store\rangle
         \xrightarrow{p}
       \langle\internaloutcome{\evalctxof{e'}},\store',\trace\rangle}
    \and
    \inferrule*[Lab=S-Emit]
      {\langle e,\store\rangle
         \estep{p}
         \langle\emitoutcome{a}{[\cdot]},\store',\trace\rangle}
      {\langle\evalctxof{e},\store\rangle
         \xrightarrow{p}
       \langle\emitoutcome{a}{E},\store',\trace\rangle}
    \and
    \inferrule*[Lab=S-Ret]
      { }
      {\langle v,\store\rangle
         \xrightarrow{p}
       \langle\emitoutcome{\actionret{v}}{[\cdot]},\store,\epsilon\rangle}
    \and
    \inferrule*[Lab=S-Abort]
      { }
      {\langle\evalctxof{\abort},\store\rangle
         \xrightarrow{p}
       \langle\emitoutcome{\actionabort}{E},\store,\epsilon\rangle}
  \end{mathpar}
  \caption{Operational semantics of honest code.}
  \label{fig:core-honest-semantics}
\end{figure}

Since expressions
are in A-normal form, an evaluation context $E$ is either a hole or a let
whose bound expression is in focus, so $E$ serves naturally as a call frame.
The store $\store$ is the run-time state.
The judgment
\smash{$\langle e,\store\rangle\estep{p}\langle
N,\store',\trace\rangle$}
steps $e$ under principal $p$ to an outcome $N$ with a trace
$\trace$ of at most one event.
The outcome either continues as an expression $e'$ or yields an action $a$
for the linked semantics together with an evaluation context.
The judgment $\xrightarrow{p}$ lifts $\estep{p}$ through evaluation
contexts, where \ruleof{S-Internal} covers internal steps and the
remaining rules cover interactions.

Most rules for the honest program's operational semantics are standard.  \ruleof{R-Op},
\ruleof{R-If} and \ruleof{R-Let} reduce operators, branches and lets as
usual, and \ruleof{R-Down} leaves its value unchanged since a downgrade
affects only the static type.
\ruleof{R-Read} and \ruleof{R-Write} access only the running
principal's cell $(p,h)$, and as is standard we treat each write as an
effect and record its update as an event.  The event shows
$\opaque{v}$, with $\opaque{\funid{q}{g}}=\unit$ and $\opaque{v}=v$ otherwise,
since a pointer is observable only by calling it.

Calls, returns and aborts are the only ways to interact with other
principals.  \ruleof{R-Call} turns a call into a $\call$ action, and
\ruleof{S-Emit} passes this action to the linked semantics along with its
call frame $E$, so the caller resumes as $E[v]$ once the call returns
$v$.  A body that reduces to a value hands it back to its caller through
\ruleof{S-Ret}, and an $\abort$ anywhere in the body ends the whole run
through \ruleof{S-Abort}.

\paragraph{Linked Semantics} Honest code and the attacker machine $\mathcal M$ are each an open
system whose interactions with the other side remain unresolved.  The linked
semantics accounts for the interaction of two heterogeneous
parties~\cite{koenig2021compcerto}, and closes the system by running both over one
shared call stack
\[
  \phi ::= \tohonest(\funid{p}{f},e)\mid\toattacker(\funid{p}{f},F)
  \qquad
  S ::= \emptyset\mid\phi::S
\]
The call stack's head is the running frame, and its tail holds the awaited callers.
An honest frame holds the expression that $\funid{p}{f}$ evaluates, and an attacker
frame holds an abstract frame $F$ whose form depends on the attacker; for the state machine $\mathcal M$,
which keeps its continuation in its state, $F$ is trivially the unit value and we
write $\toattacker(\funid{p}{f})$ for convenience.  We write
$\phi.\mathit{fn}$ for the function that a frame $\phi$ runs.

A configuration
$C=\langle s_{\calA},\store,S\rangle$ holds the attacker state $s_{\calA}$, the store
$\store$ and the stack $S$.  A run halts on an empty stack.  Each world $\mathrm W$ defines a step relation
$C\linkstep_{\mathrm W}\langle C',\trace\rangle$ that emits a trace
$\trace$ of at most one event, and its reflexive, transitive closure
$C\longrightarrow^{*}_{\mathrm W}\langle C',\trace\rangle$ concatenates the
emitted events.  Linking requires an implementation for every honest function declared in
$\sigtable$.  A run starts from an initial store $\store_{\mathrm{init}}$, with
$\mathcal M$ running the attacker-owned entry
$\funid{\top}{\entry}$ on the 
unit value.
\[
  C_{\mathrm{init}}=
  \left\langle
    \rho(s_{\calA}^{\mathrm{init}},\messageinvoke{\funid{\top}{\entry}}{\unit}),
    \store_{\mathrm{init}},
    \toattacker(\funid{\top}{\entry})::\emptyset
  \right\rangle.
\]

\begin{figure}
  \centering
  \begin{judgments}
    \judgment{
      \langle\phi,s_{\calA},\store\rangle
      \framestep_{\mathrm W}
      \langle\phi',s_{\calA}',\store',\trace\rangle}
    \and
    \judgment{
      \langle\phi,s_{\calA},\store\rangle
      \framestep_{\mathrm W}
      \langle a,s_{\calA}',\store'\rangle}
  \end{judgments}
  \begin{mathpar}[\mprset{lineskip=.8em,andskip=1.2em,sep=1em}]
    \inferrule*[Lab=F-Honest]
      {\langle\checkuse_{\mathrm W}(\funid{p}{f},e),\store\rangle\xrightarrow{p}
         \langle\internaloutcome{e'},\store',\trace\rangle}
      {\langle\tohonest(\funid{p}{f},e),s_{\calA},\store\rangle\framestep_{\mathrm W}
       \langle\tohonest(\funid{p}{f},e'),s_{\calA},\store',\trace\rangle}
    \and
    \inferrule*[Lab=F-Honest-Act]
      {\langle\checkuse_{\mathrm W}(\funid{p}{f},e),\store\rangle\xrightarrow{p}
         \langle\emitoutcome{a}{E},\store',\epsilon\rangle}
      {\langle\tohonest(\funid{p}{f},e),s_{\calA},\store\rangle\framestep_{\mathrm W}
       \langle a,s_{\calA},\store'\rangle}
    \and
    \inferrule*[Lab=F-Attacker]
      {\delta(s_{\calA})=s_{\calA}'}
      {\langle\toattacker(\funid{p}{f}),s_{\calA},\store\rangle\framestep_{\mathrm W}
       \langle\toattacker(\funid{p}{f}),s_{\calA}',\store,\epsilon\rangle}
    \and
    \inferrule*[Lab=F-Attacker-Act]
      {\delta(s_{\calA})=(a,s_{\calA}')}
      {\langle\toattacker(\funid{p}{f}),s_{\calA},\store\rangle\framestep_{\mathrm W}
       \langle a,s_{\calA}',\store\rangle}
  \end{mathpar}
  \begin{judgments}
    \judgment{
      \langle s_{\calA},\store,S\rangle
      \linkstep_{\mathrm W}
      \langle s_{\calA}',\store',S',\trace\rangle}
  \end{judgments}
  \begin{mathpar}[\mprset{lineskip=.8em,andskip=1.2em,sep=1em}]
    \inferrule*[Lab=L-Step]
      {\langle\phi,s_{\calA},\store\rangle\framestep_{\mathrm W}
         \langle\phi',s_{\calA}',\store',\trace\rangle}
      {\langle s_{\calA},\store,\phi::S\rangle\linkstep_{\mathrm W}
       \langle s_{\calA}',\store',\phi'::S,\trace\rangle}
    \and
    \inferrule*[Lab=L-Call]
      {\phi.\mathit{fn}=\funid{p}{f}
       \\
       \langle\phi,s_{\calA},\store\rangle\framestep_{\mathrm W}
         \langle\actioncall{v_f}{\ell}{v},s_{\calA}',\store'\rangle
       \\\\
       \linkcall_{\mathrm W}(s_{\calA}',\phi::S,p\join\ell,v_f,v)
         =\langle s_{\calA}'',S',\trace\rangle}
      {\langle s_{\calA},\store,\phi::S\rangle\linkstep_{\mathrm W}
       \langle s_{\calA}'',\store',S',\trace\rangle}
    \and
    \inferrule*[Lab=L-Ret]
      {\langle\phi,s_{\calA},\store\rangle\framestep_{\mathrm W}
         \langle\actionret{v},s_{\calA}',\store'\rangle
       \\\\
       \linkret_{\mathrm W}(s_{\calA}',\phi_1,S,\phi.\mathit{fn},v)=\langle s_{\calA}'',S',\trace\rangle}
      {\langle s_{\calA},\store,\phi::\phi_1::S\rangle\linkstep_{\mathrm W}
       \langle s_{\calA}'',\store',S',\trace\rangle}
    \and
    \inferrule*[Lab=L-Done]
      {\langle\phi,s_{\calA},\store\rangle\framestep_{\mathrm W}
         \langle\actionret{v},s_{\calA}',\store'\rangle}
      {\langle s_{\calA},\store,\phi::\emptyset\rangle\linkstep_{\mathrm W}
       \langle s_{\calA}',\store',\emptyset,\eventdone\rangle}
    \and
    \inferrule*[Lab=L-Abort]
      {\langle\phi,s_{\calA},\store\rangle\framestep_{\mathrm W}
         \langle\actionabort,s_{\calA}',\store'\rangle}
      {\langle s_{\calA},\store,\phi::S\rangle\linkstep_{\mathrm W}
       \langle s_{\calA}',\store',\emptyset,\eventabort\rangle}
  \end{mathpar}
  \caption{Frame and linked steps.}
  \label{fig:core-linked-semantics}
\end{figure}

\Cref{fig:core-linked-semantics} defines the linked semantics in two
layers.  The frame step $\framestep_{\mathrm W}$ advances the top frame
in isolation and produces either an updated frame or an action
$a\in\Action$.  An honest frame running $\funid{p}{f}$ steps its expression
under the honest semantics at principal $p$.  Specifically, an internal step updates the
frame (\ruleof{F-Honest}), and an action leaves the frame in place to await
a return (\ruleof{F-Honest-Act}), both of which may modify the store $\Sigma$.  Similarly, an attacker frame steps by $\delta$ to a new state
(\ruleof{F-Attacker}) or an action (\ruleof{F-Attacker-Act}).

The linked step $\linkstep_{\mathrm W}$ advances the whole configuration
according to the outcome of $\framestep_{\mathrm W}$.  \ruleof{L-Step}
 replaces the top frame with the result of one internal step.
\ruleof{L-Call} requires the called value $v_f$ to be a pointer
declared in the linked signature table
$\sigtable\bowtie\sigtable_{\calA}$. The rule then calls $v_f$ at
pc $p\join\ell$, which implies that no call is more trusted than its caller.
\ruleof{L-Ret} pops the returning frame and passes the result $v$ to
$\linkret_{\mathrm W}$ together with the caller's frame $\phi_1$,
as $v$ is treated differently for honest and adversarial
callers, as indicated below.

The helper functions $\linkcall_{\mathrm W}$ and $\linkret_{\mathrm W}$
share the same structure.  The former pushes a frame for the callee
$\funid{q}{g}$ and records $\eventcall{\funid{q}{g}}{\opaque{\pc}}$, and the latter
resumes the caller and records $\trace_p$.
The events record only what the trusted party sees~\cite{am11}, namely
the activations of honest code, with each call event marking an honest
frame's entry and each return its exit.  Calls into the attacker and
returns from it are therefore silent, and $\opaque{\pc}$ replaces an
attacker label with $\top$.

The attacker accepts every value without checks.
On a message, either
$\messageinvoke{\funid{q}{g}}{v}$ or $\messageresult{v}$, the attacker steps
into a new state via $\rho$.  In contrast, a value can be passed into
the honest party only after it has been vetted by the run-time check
in that world.
A call must pass $\checkallow_{\mathrm W}$, and $\checkadmit_{\mathrm
  W}$ converts the argument at the callee's input type $\tau_i$ and the result
at its output type $\tau_o$.  If any check fails, the run aborts.

\begin{figure}
  \centering
  \[
    \linkcall_{\mathrm W}(s_{\calA},S,\pc,v_f,v)=
    \begin{cases}
      \langle\rho(s_{\calA},\messageinvoke{\funid{q}{g}}{v}),\toattacker(\funid{q}{g})::S,
        \eventempty\rangle
        & \text{if } q\in\adv\\
      \langle s_{\calA},\tohonest(\funid{q}{g},\subst{e}{x}{v'})::S,\eventcall{\funid{q}{g}}{\opaque{\pc}}\rangle
        & \text{if } \checkallow_{\mathrm W}(S,\pc,\funid{q}{g},\pcex),\\
        & \quad\checkadmit_{\mathrm W}(\tau_i,v)=v'\\
        & \quad\text{and } \bodytable(\funid{q}{g})=\lam{x}{e}\\
      \langle s_{\calA},\emptyset,\eventabort\rangle & \text{otherwise}
    \end{cases}
  \]
  \[
    \text{where } v_f=\funid{q}{g} \text{ and }
    (\sigtable\bowtie\sigtable_{\calA})(\funid{q}{g})=
      \funsig{\tau_i}{\tau_o}{\pcex}{\pctop}
  \]
  \[
    \linkret_{\mathrm W}(s_{\calA},\phi,S,\funid{p}{f},v)=
    \begin{cases}
      \langle\rho(s_{\calA},\messageresult{v}),\toattacker(\funid{q}{g})::S,\trace_p\rangle
        & \text{if } \phi=\toattacker(\funid{q}{g})\\
      \langle s_{\calA},\tohonest(\funid{q}{g},E[v'])::S,\trace_p\rangle
        & \text{if } \phi=\tohonest(\funid{q}{g},E[\pcapp{v_f}{u}{\ell}])\\
        & \quad\text{and } \checkadmit_{\mathrm W}(\tau_o,v)=v'\\
      \langle s_{\calA},\emptyset,\eventabort\rangle & \text{otherwise}
    \end{cases}
  \]
  \[
    \text{where } (\sigtable\bowtie\sigtable_{\calA})(\funid{p}{f})=
      \funsig{\tau_i}{\tau_o}{\pcex}{\pctop}
    \text{ and }
    \trace_p=
    \begin{cases}
      \eventempty & \text{if } p\in\adv\\
      \eventret{p} & \text{otherwise}
    \end{cases}
  \]
  \caption{Call and return semantics. 
  The function $\checkallow_{\mathrm W}$ defines the run-time checks.}
  \label{fig:core-call-return}
\end{figure}

\paragraph{Dynamic Checks}
\Cref{fig:core-call-return} shows the semantics of call and return.
The dynamic checks are defined in
$\checkallow_{\mathrm W}$.  The partial function
$\checkadmit_{\mathrm W}(\tau,v)$ converts a value $v$ entering honest
code at type $\tau$ and is undefined on values the world intends to
reject.  The
total function $\checkuse_{\mathrm W}(\funid{p}{f},e)$ converts the
values that the next step of $e$ uses, before \ruleof{F-Honest} and
\ruleof{F-Honest-Act} take that step.  For instance, World 2
permits every call and passes every value unchanged:
\[
  \checkallow_{\mathrm W_2}(S,\pc,\funid{q}{g},\pcex)
  \quad\text{always holds}
  \qquad
  \checkadmit_{\mathrm W_2}(\tau,v)=v
  \qquad
  \checkuse_{\mathrm W_2}(\funid{p}{f},e)=e
\]
World 1 also passes every value unchanged, but checks each call
at run time:
\[
  \checkallow_{\mathrm W_1}(S,\pc,\funid{q}{g},\pcex)
  \quad\text{iff}\quad
  \pc\flowsto\pcex
  \land
  \forall L\in\mathit{locks}(S).\;\pc\flowsto q\join L
\]
where $\mathit{locks}(S)$ is the set of principals with an honest frame
on $S$.  The condition $\pc\flowsto\pcex$ enforces the callee's external
pc, and $\pc\flowsto q\join L$ enforces the reentrancy
lock that each honest principal $L$ on the stack holds
until it returns.  Intuitively, an attacker pc lies in $\adv$ and, by
primality, never flows to $q\join L$ for honest $q$ and $L$, so the
attacker cannot call honest code at all once an honest frame is on the
stack.

\section{Type-Confusion Security via Simulation}
\label{sec:security}

We now show that the type-confusion
security definition from \cref{sec:formal-theorems}
holds for worlds $\mathrm W_0$ and $\mathrm W_3$.

The behavior of a linked system is what an observer sees over an entire
run.  Since runs are deterministic, the traces reachable from the
initial configuration $C_{\mathrm{init}}$ form a chain under the prefix order, and the
behavior is their least upper bound, a (possibly infinite) trace.  When linking a trusted open
program $\calP$ with an attacker $\calA$ succeeds---namely, when every
honest function declared in $\sigtable$ is defined---the behavior of
running the linked system in world $\mathrm W$ from store $\store$ is:
\[
  \llbracket\calP\bowtie\calA\rrbracket_{\mathrm W}(\store)
  =
  \bigsqcup\{\,\trace \mid
    C_{\mathrm{init}}\longrightarrow^{*}_{\mathrm W}\langle C,\trace\rangle\,\},
\]
Note that a run that halts ends its trace with $\eventdone$ or $\eventabort$.
Equivalently, we can describe the behavior as the set of finite prefixes of traces.

We compare behaviors only from initial stores whose trusted cells hold values of
their declared types under IFC typing,
since the trusted program is IFC-typed.  A store is well typed for a
cell table $\celltable$ when every cell declared in $\celltable$ holds a
value of its type,
\[
  \celltable\vdash\store
  \quad\text{iff}\quad
  \forall p,h,\pc,\pctop.\;
    \celltable(p,h)=\tau
    \implies
    \ctyping{\emptyset}{\pc}{p}{\pctop}{\store(p,h)}{\tau}
\]
and we write $\vdash\store$ for $\celltable|_{\notadv}\vdash\store$, the
restriction of $\celltable$ to the principals outside $\adv$.

It remains to refine type-confusion security.

\begin{definition}[Simulation]
  \label{def:simulation}
  Consider worlds $\mathrm W$ and $\mathrm W'$ and a trusted program
  $\calP$ with $\vdash_{\mathrm W_2}\calP$.  World $\mathrm W'$ simulates
  $\mathrm W$, written
  $\mathrm W\simulates\mathrm W'$, meaning that
  $\mathrm W$'s attackers are no stronger than $\mathrm W'$'s, iff
  \[
    \forall\calA_{\mathrm W}.\;
    \vdash_{\mathrm W}\calA_{\mathrm W}
    \implies
    \exists\calA_{\mathrm W'}.\;
    \vdash_{\mathrm W'}\calA_{\mathrm W'}
    \land
    \forall\store.\;
    \vdash\store
    \implies
    \llbracket\calP\bowtie\calA_{\mathrm W}\rrbracket_{\mathrm W}(\store)
    =\llbracket\calP\bowtie\calA_{\mathrm W'}\rrbracket_{\mathrm W'}(\store)
  \]
\end{definition}

\begin{theorem}[Type-Confusion Security]
  \label{thm:security}
  If $\vdash_{\mathrm W_2}\calP$, then 
    $\mathrm W_0\approx\mathrm W_1\approx\mathrm W_2\approx\mathrm W_3$. 
    That is, \Cref{fig:worlds-overview} holds.
\end{theorem}

\subsection{Simulation between Worlds 1 and 2}
\label{sec:real-ideal}

Worlds 1 and 2 are equivalent for IFC-typed programs.  World 1's
weaker type system lets attackers confuse IFC labels, and
the run-time check removes exactly this power.

The direction $\mathrm W_2\simulates\mathrm W_1$ is almost trivial.
Consider an
attacker $\calA_{\mathrm W_2}=(\adv,\sigtable_{\calA},\mathcal M)$ that
is boundary-typed in $\mathrm W_2$.
Against an IFC-typed program,
$\calA_{\mathrm W_2}$ emits only actions well typed in $\mathrm W_2$, which
pass the run-time check, so $\calA_{\mathrm W_2}$ behaves the same in
both worlds.
The only complication arises when reasoning about $\calA_{\mathrm
  W_2}$'s boundary
typing in $\mathrm W_1$, which also covers messages that are typed in
$\mathrm W_1$ but not in $\mathrm W_2$, on which $\calA_{\mathrm W_2}$ may
emit an action ill typed in $\mathrm W_1$.  The simulator
$\calA_{\mathrm W_1}$ therefore runs $\mathcal M$ behind a guard.  Its
transition function $\delta_{\mathrm W_1}$ follows $\delta$, except that an
action $a$ with $\funid{p}{f}\nvdash_{\mathrm W_1}a$ is replaced by an
internal step that leaves the state unchanged, where $\funid{p}{f}$ is
the running function.  Against an IFC-typed program,
$\delta_{\mathrm W_1}$ coincides with $\delta$.

For $\mathrm W_1\simulates\mathrm W_2$, the simulator
runs $\calA_{\mathrm W_1}$ unchanged in $\mathrm W_2$ since such an attacker
respects ordinary types but not labels.
Moreover, both the functions a $\mathrm W_1$ attacker passes to honest code and the calls it
makes may carry labels that IFC typing rejects.

\begin{figure}
  \centering
  \tikzset{
    box/.style={fill=black!8,rounded corners},
    call/.style={->,>=stealth},
    int/.style={call,dash pattern=on 2pt off 1.5pt},
    lbl/.style={font=\scriptsize,inner sep=1pt},
    world/.style={font=\scriptsize\bfseries},
  }
  \begin{subfigure}[b]{0.3\textwidth}
    \centering
    \begin{tikzpicture}
      \path (0,1.1) -- (0,-2.4);
      \node[world] at (-0.5,0) {$\mathrm W_1$};
      \path[box] (0,-0.45) rectangle (2,0.45);
      \node[above] at (1,0.45) {$\calA_{\mathrm W_1}$};
      \node (f1) at (0.3,0) {$f'$};
      \node (f) at (1.7,0) {$f$};
      \draw[call] ([yshift=3pt]f1.east) -- node[lbl,above] {$\mathsf{call}$} ([yshift=3pt]f.west);
      \draw[call] ([yshift=-3pt]f.west) -- node[lbl,below] {$\mathsf{ret}$} ([yshift=-3pt]f1.east);
      \node[world] at (-0.5,-1.9) {$\mathrm W_2$};
      \path[box] (0,-2.35) rectangle (2,-1.45);
      \node[above] at (1,-1.45) {$\calA_{\mathrm W_2}$};
      \node (f1') at (0.3,-1.9) {$f'$};
      \node (f') at (1.7,-1.9) {$f$};
      \draw[int] ([yshift=3pt]f1'.east) -- node[lbl,above] {$\mathsf{call}$} ([yshift=3pt]f'.west);
      \draw[int] ([yshift=-3pt]f'.west) -- node[lbl,below] {$\mathsf{ret}$} ([yshift=-3pt]f1'.east);
    \end{tikzpicture}
    \caption{Attacker to Attacker}
    \label{fig:real-ideal-internal}
  \end{subfigure}
  \hfill
  \begin{subfigure}[b]{0.3\textwidth}
    \centering
    \begin{tikzpicture}
      \path (0,1.1) -- (0,-2.4);
      \node[world] at (-0.5,0) {$\mathrm W_1$};
      \node (g) at (0,0) {$g$};
      \path[box] (1.3,-0.45) rectangle (2.1,0.45);
      \node[above] (A) at (1.7,0.45) {$\calA_{\mathrm W_1}$};
      \node (f) at (1.7,0) {$f$};
      \draw[call] (A.west) to[bend right=30] node[lbl,above] {$\funid{r}{f}$} (g.north);
      \draw[call] ([yshift=3pt]g.east) -- node[lbl,above] {$\mathsf{call}$} ([yshift=3pt]f.west);
      \draw[call] ([yshift=-3pt]f.west) -- node[lbl,below] {$\mathsf{ret}$} ([yshift=-3pt]g.east);
      \node[world] at (-0.5,-1.9) {$\mathrm W_2$};
      \node (g') at (0,-1.9) {$g$};
      \path[box] (1.3,-2.35) rectangle (3.5,-1.45);
      \node[above] (A') at (2.4,-1.45) {$\calA_{\mathrm W_2}$};
      \node (ad) at (1.7,-1.9) {$f^{\tau}$};
      \node (f') at (3.2,-1.9) {$f$};
      \draw[call] (A'.west) to[bend right=30] node[lbl,above] {$\funid{p}{f^{\tau}}$} (g'.north);
      \draw[call] ([yshift=3pt]g'.east) -- node[lbl,above] {$\mathsf{call}$} ([yshift=3pt]ad.west);
      \draw[call] ([yshift=-3pt]ad.west) -- node[lbl,below] {$\mathsf{ret}$} ([yshift=-3pt]g'.east);
      \draw[int] ([yshift=3pt]ad.east) -- node[lbl,above] {$\mathsf{call}$} ([yshift=3pt]f'.west);
      \draw[int] ([yshift=-3pt]f'.west) -- node[lbl,below] {$\mathsf{ret}$} ([yshift=-3pt]ad.east);
    \end{tikzpicture}
    \caption{Honest to Attacker}
    \label{fig:real-ideal-adapter}
  \end{subfigure}
  \hfill
  \begin{subfigure}[b]{0.3\textwidth}
    \centering
    \begin{tikzpicture}
      \path (0,1.1) -- (0,-2.4);
      \node[world] at (-0.5,0) {$\mathrm W_1$};
      \path[box] (0,-0.45) rectangle (0.7,0.45);
      \node[above] at (0.35,0.45) {$\calA_{\mathrm W_1}$};
      \node (f) at (0.35,0) {$f$};
      \node (g) at (3.3,0) {$g$};
      \draw[call] ([yshift=3pt]f.east) -- node[lbl,above] {$\mathsf{call}^{\pcex}$} ([yshift=3pt]g.west);
      \draw[call] ([yshift=-3pt]g.west) -- node[lbl,below] {$\mathsf{ret}$} ([yshift=-3pt]f.east);
      \node[world] at (-0.5,-1.9) {$\mathrm W_2$};
      \path[box] (0,-2.35) rectangle (2.3,-1.45);
      \node[above] at (1.2,-1.45) {$\calA_{\mathrm W_2}$};
      \node (f') at (0.35,-1.9) {$f$};
      \node (tr) at (1.9,-1.9) {$\mathit{tr}_{\pcex}$};
      \node (g') at (3.3,-1.9) {$g$};
      \draw[call] ([yshift=3pt]f'.east) -- node[lbl,above] {$\mathsf{call}$} ([yshift=3pt]tr.west);
      \draw[call] ([yshift=-3pt]tr.west) -- node[lbl,below] {$\mathsf{ret}$} ([yshift=-3pt]f'.east);
      \draw[call] ([yshift=3pt]tr.east) -- node[lbl,above] {$\mathsf{call}^{\pcex}$} ([yshift=3pt]g'.west);
      \draw[call] ([yshift=-3pt]g'.west) -- node[lbl,below] {$\mathsf{ret}$} ([yshift=-3pt]tr.east);
    \end{tikzpicture}
    \caption{Attacker to Honest}
    \label{fig:real-ideal-trampoline}
  \end{subfigure}
  \captionsetup{font=footnotesize}
  \caption{Simulator $\calA_{\mathrm W_2}$ for $\calA_{\mathrm W_1}$.  
  Shaded boxes contain attacker functions; $\g$ is honest. 
  Dashed arrows denote internal simulation steps.
  }
  \label{fig:real-ideal-simulator}
\end{figure}
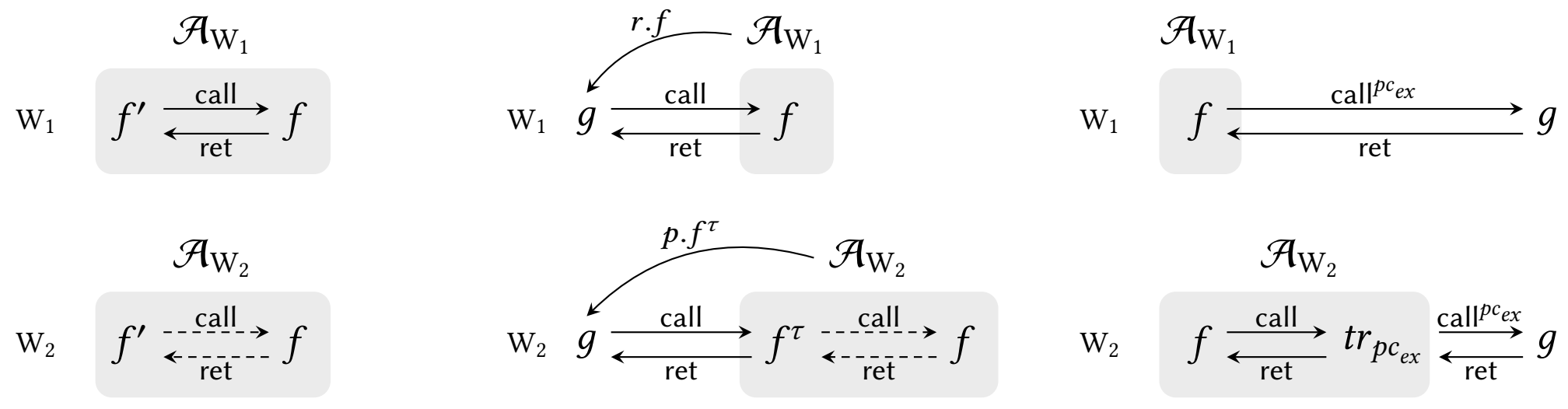

\Cref{fig:real-ideal-simulator} sketches the simulator $\calA_{\mathrm W_2}$
for $\calA_{\mathrm W_1}$. 
 First, a call between
two functions owned by $\calA_{\mathrm W_1}$
emits no observable event, but the labels of the passed arguments may
not be IFC-typed, which violates boundary typing.  $\calA_{\mathrm W_2}$
therefore emulates such a call in-house
(\cref{fig:real-ideal-internal}) instead of explicitly emitting such a call.

Then, $\calA_{\mathrm W_1}$ may hand a pointer to one of its functions
$f$ to honest code, which expects a type
$\tau=\funptrtype{\tau_i}{\tau_o}{\pcex}{\pctop}{\ell}$. Ordinary typing, however, checks $f$ only against
$\erase{\tau}$, so a label mismatch may leave $\emptyset;\pc;p';\pctop\nvdash_{\mathrm W_2}f:\tau$
for all allowed labels.
In this case, $\calA_{\mathrm W_2}$ instead hands over a pointer to an adapter $f^{\tau}$
(\cref{fig:real-ideal-adapter}), a fresh function whose signature is derived from $\tau$.
Specifically, a pointer to $f^{\tau}$ has type
$\funptrtype{\tau_i\join\top}{\tau_o}{\pcex\join p}{\pctop\meet p}{p}$,
which is a subtype of $\tau$ at label $p$.  Since a pointer is labeled by the owner of its
function, $\calA_{\mathrm W_2}$ picks $p=\ell\join q$, which is an
attacker principal by upward closure of $\adv$.  Here $q$ is the owner
of the function of $\calA_{\mathrm W_1}$ that returns the pointer, because
a returned value also carries the label of its function, and $q=\bot$
when the pointer is passed as an argument instead.
The adapter has no behavior of its own; when honest code calls
$f^{\tau}$, $\calA_{\mathrm W_2}$ runs $\calA_{\mathrm W_1}$ at $f$.  Honest
code may also hand this pointer back, and $\calA_{\mathrm W_2}$ then
replaces it by the pointer to $f$, which is what $\calA_{\mathrm W_1}$ would receive in
$\mathrm W_1$.

Finally, $\calA_{\mathrm W_1}$ may call an honest function $g$ with external
pc $\pcex$, which the run-time check of $\mathrm W_1$
(\cref{sec:dynamic-semantics}) permits only when no honest frame is on
the stack.  IFC typing, however, rejects this call from the outermost frame of
$\calA_{\mathrm W_2}$, which runs at $\top$.
$\calA_{\mathrm W_2}$ therefore detours through a trampoline
$\mathit{tr}_{\pcex}$ (\cref{fig:real-ideal-trampoline}), a fresh function declared at
$\funsig{\tunit}{\tunit}{\top}{\bot}$, which any frame may call.  The
trampoline then completes the call of $g$ at $\pcex$, which is an attacker principal
due to upward
closure of $\adv$.  Because $\mathit{tr}_{\pcex}$ serves every
call with external pc $\pcex$, it takes only a unit argument, so the
actual argument travels through the private state of
$\calA_{\mathrm W_2}$, with each function
pointer replaced by its adapter at $\tau_i$, and the result returns the
same way.  The program mentions only finitely many types and
labels, so $\calA_{\mathrm W_2}$ declares only finitely many adapters and
trampolines.

 $\calA_{\mathrm W_2}$ must abort whenever $\mathrm W_1$ refuses a
 call. This happens when a call of $\calA_{\mathrm W_1}$ into honest code
fails the run-time check, and when honest code calls a pointer to an
honest function that $\calA_{\mathrm W_1}$ handed over.  That pointer
carries an attacker label, so in $\mathrm W_1$ honest code calls it at an
attacker pc above its own frame, which the run-time check refuses,
whereas in $\mathrm W_2$ the call reaches an adapter.  As in the proof
of $\mathrm W_2\simulates\mathrm W_1$, a guard for $\mathrm W_2$ makes
$\calA_{\mathrm W_2}$ boundary-typed by construction.

The proof relates each configuration of $\mathrm W_1$ running
$\calA_{\mathrm W_1}$ to one of $\mathrm W_2$ running $\calA_{\mathrm W_2}$,
where both hold the same state of $\calA_{\mathrm W_1}$ and their honest
code and cells agree once adapters are renamed back.  Each step in World 1 is
matched by one step in World 2, or by two through a trampoline, where the
extra step is silent since the trampoline belongs to the attacker.  When $\mathrm W_1$ is stuck, $\calA_{\mathrm W_2}$ spins
silently, and on related runs the guard never intervenes. Therefore,
both runs
produce the same behavior, and we can conclude that $\mathrm W_1\approx\mathrm W_2$.

\subsection{Simulation between Worlds 0 and 1}
\label{sec:base-real}
In reality, an attacker need not respect even ordinary types.  It
may hand honest code a number or a pointer where a boolean
is expected, and the honest code must somehow interpret that
value. This interpretation is modeled by a function $\cast$, where $\cast(v,\sigma)$ is $v$
reinterpreted at an unlabeled type $\sigma$.  We assume that a cast to a data type
$\realbooltype$, $\realnattype$ or $\tunit$ yields a value of that
type, while a cast to a pointer type may not.  Honest code normalizes such a value
using $\mathit{coerce}(v,\sigma)$, which returns $v$ if it has type $\sigma$,
and $\cast(v,\sigma)$ otherwise.  The result is therefore well typed at
every data type.


Like World 1, World 0 changes only the run-time checks
(\cref{sec:dynamic-semantics}).  It keeps the lock check,\linebreak
\mbox{$\checkallow_{\mathrm W_0}=\checkallow_{\mathrm W_1}$}, and applies
$\mathit{coerce}$ to pointers and data at different points.  A
pointer is coerced eagerly as it crosses the boundary.  If
$\sigma=\erase{\tau}$ is not a data type, then $\checkadmit_{\mathrm
  W_0}(\tau,v)$ is $\mathit{coerce}(v,\sigma)$ when that value has type
$\sigma$, and the run aborts otherwise.  When $\sigma$ is a data type,
$v$ passes unchanged.  Intuitively, since the
signature table is fixed at compile time, an ill-typed pointer cannot
be repaired, so honest code must reject it before running.

Data is instead coerced lazily.  The check $\checkuse_{\mathrm W_0}$ coerces such a value at
the type of that use and leaves every other expression unchanged,
\[
  \begin{array}{@{}r@{\;}l@{}}
    \checkuse_{\mathrm W_0}(\funid{p}{f},E[\exprif{v}{e_1}{e_2}])
      &= E[\exprif{\mathit{coerce}(v,\realbooltype)}{e_1}{e_2}]\\
    \checkuse_{\mathrm W_0}(\funid{p}{f},E[\exprop{v_1}{v_2}])
      &= E[\exprop{\mathit{coerce}(v_1,\sigma_{\op})}{\mathit{coerce}(v_2,\sigma_{\op})}]\\
    \checkuse_{\mathrm W_0}(\funid{p}{f},E[\writeexpr{h}{v}])
      &= E[\writeexpr{h}{\mathit{coerce}(v,\erase{\celltable}(p,h))}]\\
    \checkuse_{\mathrm W_0}(\funid{p}{f},E[\pcapp{\funid{q}{g}}{v}{\ell}])
      &= E[\pcapp{\funid{q}{g}}{\mathit{coerce}(v,\sigma_i)}{\ell}]\\
    \checkuse_{\mathrm W_0}(\funid{p}{f},v)
      &= \mathit{coerce}(v,\sigma_o')
  \end{array}
\]
where $\sigma_{\op}$ is the operand type of $\op$, $\sigma_i$ is the
input type of $\erase{\sigtable}(\funid{q}{g})$ and $\sigma_o'$ is
the output type of $\erase{\sigtable}(\funid{p}{f})$. A
return also counts as a use, so a confused value never leaves honest
code.

As in \cref{sec:real-ideal}, we prove $\mathrm W_0\approx\mathrm W_1$ by
showing that each world simulates the other
(\cref{def:simulation}).  The direction $\mathrm W_1\simulates\mathrm W_0$
is immediate.  An attacker boundary-typed in $\mathrm W_1$ is also an
attacker in $\mathrm W_0$.  By the boundary typing of $\calA_{\mathrm W_1}$, every value
that reaches honest code is well typed. As a consequence, $\checkadmit_{\mathrm W_0}$
and $\checkuse_{\mathrm W_0}$ do not change values, and the two worlds run in lockstep.

Conversely, an attacker $\calA_{\mathrm W_0}$ of $\mathrm W_0$ may send
ill-typed values, which no boundary-typed attacker in $\mathrm W_1$ can.
As a result, we construct the
simulator $\calA_{\mathrm W_1}$ to run $\calA_{\mathrm W_0}$ and cast
eagerly.  The simulator replaces each value $v$ that
$\calA_{\mathrm W_0}$ sends to honest code at type
$\sigma$ by the value that $\mathrm W_0$
eventually uses (i.e., $\mathit{coerce}(v,\sigma)$), and aborts whenever $\mathrm W_0$ rejects an
ill-typed pointer.

A call of
$\calA_{\mathrm W_0}$ to its own function $h$ raises a further problem, since
$\calA_{\mathrm W_0}$ may pass $h$ an argument of any type, which
$\mathrm W_0$ hands to $h$ unchanged.  Coercing it would change what $h$
receives, so, as in the proof
of $\mathrm W_1\approx\mathrm W_2$ (\cref{sec:real-ideal}), $\calA_{\mathrm W_1}$ instead calls a
fresh trampoline of $h$, declared at
$\realfunctiontype{\tunit}{\tunit}$, with $\unit$, and the actual
argument and result travel through the private state of
$\calA_{\mathrm W_1}$.  Every action of
$\calA_{\mathrm W_1}$ is therefore boundary-typed in $\mathrm W_1$, and the
two worlds run in lockstep, so $\mathrm W_0\approx\mathrm W_1$.

\subsection{Simulation between Worlds 2 and 3}
\label{sec:ideal-code}

An attacker in World 2 is an arbitrary state machine, whose state and
transitions need not be expressible in our language.  We show that
this generality gives the attacker no additional power by expressing
every World 2 attacker as well-typed code with stateless
external functions.

\paragraph{World 3 Attacker} Attacker code
$\calA=(\adv,\coreiface_{\calA},\bodytable_{\calA},X_{\calA},\store_{\calA})$
consists of an interface
$\coreiface_{\calA}=(\sigtable_{\calA},\celltable_{\calA})$ at
principals in $\adv$, a body table $\bodytable_{\calA}$, a table
$X_{\calA}:\bbP\times\FunName\finpartial(\Value\to\Value)$ of stateless external
functions private to the attacker, and a store $\store_{\calA}$.
Unlike a state machine, whose frame $F$ in
$\toattacker(\funid{p}{f},F)$ is trivial
(\cref{sec:dynamic-semantics}), attacker code follows honest code
and takes $F$ to be the expression $e$ that $\funid{p}{f}$ evaluates.
World 3 keeps the linked semantics and changes only the rules
for attacker frames, which now step by the operational semantics of
the honest party on $\store_{\calA}$ without emitting any events. That
is, we replace \ruleof{F-Attacker}
and \ruleof{F-Attacker-Act} with the
following rules:
\begin{mathpar}
  \inferrule*[Lab=F-Code]
    {\langle e,\store_{\calA}\rangle\xrightarrow{p}
       \langle\internaloutcome{e'},\store_{\calA}',\trace\rangle}
    {\langle\toattacker(\funid{p}{f},e),\store_{\calA},\store\rangle\framestep_{\mathrm W_3}
     \langle\toattacker(\funid{p}{f},e'),\store_{\calA}',\store,\epsilon\rangle}
  \and
  \inferrule*[Lab=F-Code-Act]
    {\langle e,\store_{\calA}\rangle\xrightarrow{p}
       \langle\emitoutcome{a}{E},\store_{\calA}',\epsilon\rangle}
    {\langle\toattacker(\funid{p}{f},e),\store_{\calA},\store\rangle\framestep_{\mathrm W_3}
     \langle a,\store_{\calA}',\store\rangle}
\end{mathpar}
Consequently, calls into and returns to attacker code no longer pass
through $\rho$, the receive function of the state machine.  As with
honest code, a call to $\funid{q}{g}$ with $v$ pushes the frame
$\toattacker(\funid{q}{g},\subst{e}{x}{v})$ when
$\bodytable_{\calA}(\funid{q}{g})=\lam{x}{e}$, or
$\toattacker(\funid{q}{g},X_{\calA}(\funid{q}{g})(v))$ when
$\funid{q}{g}$ is external, and a result returned to an attacker frame
is plugged into its suspended expression.

Attacker code is well typed when its interface agrees with the
program's, its bodies and cells are well typed against the linked
interface, and its external functions are private and respect their
signatures,
\[
\begin{aligned}
  \vdash_{\mathrm W_3}\calA
  \quad\Longleftrightarrow\quad
  &\coreiface\asymp\coreiface_{\calA}
  \land\coreiface\bowtie\coreiface_{\calA}\vdash_{\mathrm W_2}\bodytable_{\calA}
  \land\celltable_{\calA}\vdash\store_{\calA}\\
  \land\;&\forall\,\funid{p}{f}\in\dom(X_{\calA}).\;
    \funid{p}{f}\notin\dom(\sigtable)\\
  \land\;&\forall\,\funid{p}{f}\in\dom(X_{\calA}),v.\;
    \vdash_{\mathrm W_2}\messageinvoke{\funid{p}{f}}{v}
    \implies
    \funid{p}{f}\vdash_{\mathrm W_2}\messageresult{X_{\calA}(\funid{p}{f})(v)}
\end{aligned}
\]
where $\coreiface\asymp\coreiface_{\calA}$ means the two interfaces
agree on every function and cell they both declare (\cref{sec:notation-overload}), and
$\coreiface\bowtie\coreiface_{\calA}$ is their componentwise link
$(\sigtable\bowtie\sigtable_{\calA},\celltable\bowtie\celltable_{\calA})$.
$\celltable\bowtie\celltable_{\calA})$ is defined the same way as linking in
\cref{sec:attackers} (\cref{sec:notation-overload}).

As before, $\mathrm W_3\approx\mathrm W_2$ follows from two simulations.
The direction $\mathrm W_3\simulates\mathrm W_2$ holds by type safety.  Given
well-typed code $\calA$, we construct a state machine simulator whose state is
$\store_{\calA}$ and the frames of $\calA$.  The simulator has the same
behaviors as $\calA$, since its receive function $\rho$ starts the body
of $\calA$ on each invocation and continues the waiting call on each
result, and its transition $\delta$ steps that call by \ruleof{F-Code}
and \ruleof{F-Code-Act}.  The simulator is boundary-typed in
$\mathrm W_2$, since by progress and preservation every action that
well-typed code emits is well typed.

The converse direction $\mathrm W_2\simulates\mathrm W_3$ is harder, since
$\mathcal M$ may be any state machine, while attacker code has only
typed cells and stateless external functions.  Given an attacker
$\calA_{\mathrm W_2}=(\adv,\sigtable_{\calA},\mathcal M)$, we construct
code $\calA_{\mathrm W_3}$ that runs $\mathcal M$ step by step.  The code
keeps the execution history of $\mathcal M$ in a single cell
$\mathit{h_\text{hist}}$ of type $\tnatof{\top}$, which records every invocation
and every result that $\mathcal M$ receives and every step that
$\mathcal M$ takes.  For each function $f$ of $\calA_{\mathrm W_2}$, the
code of $f$ in $\calA_{\mathrm W_3}$ records the invocation in
$\mathit{h_\text{hist}}$ and then repeatedly obtains the next step of
$\mathcal M$ from an external function that replays the history,
recording and performing each such step.

The code $\calA_{\mathrm W_3}$ has the same behavior as
$\calA_{\mathrm W_2}$, since $\calA_{\mathrm W_3}$ performs every step of
$\mathcal M$ with only silent bookkeeping steps in between.  The code
$\calA_{\mathrm W_3}$ is also well typed, since the external functions
of $\calA_{\mathrm W_3}$ return a default value of the declared type on
any input whose replay yields an ill-typed value, and the code of $f$
would diverge on a call to a function that $f$ may not call.  Neither
fallback occurs, since every history in $\mathit{hist}$ arises from a
run of $\mathcal M$ and $\calA_{\mathrm W_2}$ is boundary-typed.

\section{Applications}
\label{sec:applications}

We can now prove that well-typed programs are free of confused
deputy and reentrancy attacks in all worlds, $W_0$ to $W_3$.  Each
property is proved in a single world and transfers to the other worlds by
\cref{thm:security}.  

\subsection{Confused Deputy Attacks}
\label{sec:app-cda}

A CDA happens when an attacker calls a trusted
function from a context less trusted than the function's external pc
admits.
\begin{definition}[CDA Security]
  \label{def:no-cda}
  Consider a program $\calP$ and a world $\mathrm W$.
  We say that $\calP$ is free from confused-deputy attacks in $\mathrm W$
  if, for all attackers $\calA_{\mathrm W}$ and stores $\store$
  such that $\vdash_{\mathrm W}\calA_{\mathrm W}$ and $\vdash\store$ and
 $\trace = \llbracket\calP\bowtie\calA_{\mathrm W}\rrbracket_{\mathrm W}(\store)$,
 the following holds.
  \[
    \forall\eventcall{\funid{q}{g}}{\pc}\in\trace.\;
    (\pc\in\adv\implies\pcex\in\adv)\land
    (\pc\notin\adv\implies\pc\flowsto\pcex)
  \]
where 
$\sigtable(\funid{q}{g})=\funsig{\tau_i}{\tau_o}{\pcex}{\pctop}$
and $\adv$ is the set of attacker-controlled labels.
\end{definition}
For an attacker call, \cref{def:no-cda} requires only
$\pcex\in\adv$. This is because the trace shows the attacker's pc as
$\top$ and, by upward closure of $\adv$, an attacker pc that flows to
$\pcex$ places $\pcex$ in $\adv$.

\begin{theorem}[CDA Security]
  \label{thm:no-cda}
  If $\vdash_{\mathrm W_2}\calP$, then $\calP$ is free of confused-deputy
  attacks in $\mathrm W_0$ to $\mathrm W_3$.
\end{theorem}
\begin{proof}
  We prove the theorem for $W_2$, and the result for the other worlds follows from \cref{thm:security}.  In $W_2$, \ruleof{TI-App} types every call into honest code,
  including the calls of the boundary-typed attacker.  The premise
  $\pc\join\ell\flowsto\pcex$ of \ruleof{TI-App} gives
  $\pc\flowsto\pcex$.  If $\pc\notin\adv$, the second case holds.  If
  $\pc\in\adv$, then $\pcex\in\adv$ by upward closure of $\adv$, which
  proves the
  first case.
\end{proof}

\subsection{Reentrancy Attacks}
\label{sec:app-reentrancy}

A reentrancy attack, as in \cref{fig:reentrancy}, calls back into
trusted code while a trusted function waits on an untrusted call.  We
write $|\trace|_{\mathsf{call}}$ and $|\trace|_{\mathsf{ret}}$ for the
numbers of call and return events in a trace $\trace$, and
$\trace'\preceq\trace$ when $\trace'$ is a prefix of $\trace$.
\begin{definition}[Reentrancy Security]
  \label{def:no-reentrancy}
  We say that $\calP$ is free from reentrancy attacks in $\mathrm W$
  if, for all attackers $\calA_{\mathrm W}$ and stores $\store$
  such that $\vdash_{\mathrm W}\calA_{\mathrm W}$ and $\vdash\store$, 
  and $\trace = \llbracket\calP\bowtie\calA_{\mathrm W}\rrbracket_{\mathrm W}(\store)$,
  and for all traces $\trace_1$ and $\trace_2$, functions $\funid{p}{f}$ and
  $\funid{q}{g}$, and labels $\pc$ and $\pc'$,
 the following holds.
  \[
    \trace_1::\eventcall{\funid{p}{f}}{\pc'}::\trace_2::\eventcall{\funid{q}{g}}{\pc}\preceq\trace
    \land
    (\forall\trace'\preceq\trace_2.\;|\trace'|_{\mathsf{ret}}\le|\trace'|_{\mathsf{call}})
    \implies
    \pc\flowsto q\join p.
  \]
\end{definition}
The second premise states that $\funid{p}{f}$ has not returned when
$\funid{q}{g}$ is called.  Note that the attacker calls no honest code
while an honest function waits, since the trace shows the attacker's pc
as $\top$, which by primality never flows to $q\join p$.

\begin{theorem}[Reentrancy Security]
  \label{thm:no-reentrancy}
  If $\vdash_{\mathrm W_2}\calP$, then $\calP$ is free of reentrancy
  attacks in $\mathrm W_0$ to $\mathrm W_3$.
\end{theorem}
\begin{proof}
  We prove the theorem for $W_1$, and the result for the other worlds follows from \cref{thm:security}. 
  In $W_1$, when $p$ is a lock on the call stack when $\funid{q}{g}$ is called,
  the dynamic check $\checkallow_{\mathrm W_1}$ ensures that the calling pc flows to
  $q\join p$.  An honest calling pc appears unchanged in the trace, and
  an attacker pc fails the check by primality.
\end{proof}

\section{Related Work}
\label{sec:related-work}

At a high level, our proof follows
the standard context-based backtranslation strategy
\cite{10.1145/2951913.2951941,Devriese2017, 10.1145/2951913.2951941},
where the attackers are contexts.
Instead of defining a translation of calculus terms, we use a technique
standard in the tradition of state-machine refinement~\cite{LynchT87}
to separate the argument about boundary typing from the language
implementing the attacker.

\citet{riely2003trust,safedpi}
aim to provide secure execution in a
decentralized system in which not all code is well-typed.
They have a different system setting:
instead of RPC calls, principals send code blocks to each other.
Besides, their major result is subject reduction, while we define and prove
a novel security objective.

\citet{rajaniAccessControlCapabilities2016} formally characterize
CDA vulnerabilities and show that capabilities cannot prevent all CDAs.
Compared to their setting, our system
supports first-class function pointers,
and we assume a weaker runtime.
They showed that CDAs are only prevented when \emph{Full Provenance}
is enforced.
This is the same insight as using information flow control
for both control flow and data flow in our work.
Our major theoretical result is a simulation theorem that generalizes
to other kinds of attacks, and CDA security is just an instance of our results.

Our approach to reentrancy is inspired by SeRIF \cite{serif21},
a core calculus that enforces secure reentrancy using a combination
of both static and dynamic IFC.
World 1's dynamic check is similar to that of SeRIF,
but SeRIF does not handle type confusion.

The topic of our work resembles that of gradual typing \cite{gradual-typing, wf09},
but we have different settings and goals.
At a high level, gradual typing studies how static type checking
and dynamic type checking can be combined in a
complementary way;
type confusion is interested in providing assurance
when code is linked to untrusted code that is
neither statically nor dynamically typed.
As a result, the two approaches yield different theorems \cite{Siek2015RefinedCF, wf09}.
Gradual IFC languages \cite{hlio, toro18gradual, 9505218,
quest-complete} are not fully decentralized; they rely on a
system-wide runtime that enforces IFC on both the attacker and the
honest parties.

Our enforcement against ordinary-type confusion and reentrancy mirrors the idea
of Control Flow Integrity (CFI)~\cite{cfi, mcfi}, a security policy that
requires that all executions follow paths in a predetermined control-flow
graph. This current work is more general: the above attacks are 
only a subset of attacks that our framework handles.

\section{Conclusion}
\label{sec:conclusion}
Although IFC type systems provide a
well-studied approach to end-to-end security,
their adoption remains challenging.
This paper develops type-confusion security for IFC type systems, 
supporting modular enforcement of both data-flow and control-flow security.
A promising future direction is to use type confusion
as the security objective to guide the development of systems
that modularly adopt new type systems.

%


\section*{AI Use}
 Codex and Claude Code assisted with Lean~\cite{moura2021lean}
 development under human
 guidance, including assisting in the translation of pen-and-paper definitions and proofs,
 auditing proofs, and mechanizing the motivating examples. We designed
 the structure of the mechanization, including the language design,
 state-machine formulation, and the right abstractions, and
 we audited the result.
%
%
\finalpage 

\section*{Data-Availability Statement}
The supplementary material for this paper includes two artifacts: an appendix in PDF format and a Lean~4 
\cite{moura2021lean}
mechanization of the paper's results.
We will submit the Lean artifact for evaluation.

The Lean~4 mechanization includes all of the results, including the examples, from the paper as
well as a file called \texttt{README.md}, which maps each result of
the paper to its definition in the Lean artifact.
The Lean artifact also contains some results that were not extensively
discussed in the paper, such as type safety of both the honest
programs and the system. The artifact contains no \texttt{sorry} and
every result depends only on the standard classical axioms of Lean, namely propositional extensionality (\texttt{propext}),
quotient soundness (\texttt{Quot.sound}), and choice (\texttt{Classical.choice}),
which can be checked by running \texttt{make axiom}
(\texttt{AxiomsAudit.lean}).
The artifact uses Lean~4 and Mathlib~\cite{community2020lean}.

The appendix includes more examples on CDA attacks, and missing
definitions referred to by the main text.

\begingroup
  \small
  \bibliography{bibtex/pm-master, blockchain, related}
\endgroup

\clearpage
\appendix
\section{Additional Examples: Model Context Protocol and Webhost}
\label{sec:other-applications}

As we suggested in the introduction, the threat
model is not exclusive to smart contracts.
\Cref{fig:reskin} shows two additional applications that are subject to
the same attacks.

The code on the left implements an MCP (Model Context Protocol) host.
The Model Context Protocol (MCP) connects AI applications to external tools,
such as file editors and deployment services.
The threat model of this MCP host is very similar to that of a smart contract:
the actions and callbacks are often provided by third-party libraries with
no standard way for specifying permission policies,
and callbacks and callers are built using heterogeneous toolchains.
A CDA substitutes the privileged tool executor for \texttt{sendToChat}, violating
the executor's external-pc restriction and executing a tool-generated proposal
without separate approval.
A reentrant callback violates its bounding pc by invoking the host
again before the approval is marked as used, allowing one authorization to execute
the action multiple times.
A CFI attack supplies data in place of \texttt{sendToChat},
causing that data to be interpreted as a function pointer and
redirecting execution to an unintended function.
These attacks follow the same patterns as the exchange-service attacks
explained above.

A cloud website service builds source code into deployable sites.
As the service evolves, it may adopt IFC typing incrementally while
retaining callbacks from older components.
The host cannot assume that these callbacks obey its new security discipline,
even when they are maintained by the same organization.
The host in \Cref{fig:reskin} calls the client's \texttt{archiveBuild} callback
to upload the artifact to the client's repository before verifying and
publishing the site. Archiving also retains rejected builds for diagnosis.
A CDA substitutes the privileged publisher for \texttt{archiveBuild}, violating
the publisher's external-pc restriction and publishing a site before verification.
If builds share a mutable staging slot, a reentrant callback can violate its
bounding pc by starting a nested trusted build that overwrites the pending
release's artifact.
A CFI attack supplies data in place of \texttt{archiveBuild},
causing that data to be interpreted as a function pointer and
redirecting execution to an unintended function.
These attacks likewise follow the patterns of the exchange-service attacks.

\begin{figure}[t]
    \centering
    \captionsetup{skip=4pt}
    \begingroup
\lstset{
    basicstyle=\scriptsize\ttfamily,
    numberstyle=\tiny\itshape,
    literate={},
    keywordstyle=[1]\bfseries\color{darkBlue},
    keywordstyle=[2]\bfseries\color{black},
    keywordstyle=[4]\color[rgb]{0.08,0.32,0.24},
    frame=none,
    aboveskip=0pt,
    belowskip=0pt,
    morekeywords=[1]{Unit, Never, JSON, String, Artifact},
    morekeywords=[3]{mcp, approvals, tools, cloud},
    morekeywords=[4]{runApproved, isUsed, approvedAction, execute, sendToChat,
        markUsed, buildAndPublish, build, archiveBuild, verify, publish, onReject},
    deletedelim=[s]{\{}{\}},
    moredelim=[s][\bfseries\color{darkBlue}]{\{U}{\}},
    moredelim=[s][\bfseries\color{darkBlue}]{\{T}{\}}
}
\begin{tikzpicture}[
    codebox/.style={
        draw=black!65,
        fill=black!4,
        line width=0.5pt,
        rounded corners=3pt,
        inner sep=3pt,
        outer sep=0pt
    },
    boxlabel/.style={
        font=\scriptsize\sffamily,
        fill=none,
        inner xsep=2pt,
        inner ysep=1pt,
        anchor=south west,
        yshift=1pt
    }
]
    \node[codebox, anchor=north west] (reskin-mcp) at (0, 0) {%
        \begin{minipage}{0.47\linewidth}
            \vspace*{2pt}
            \lstinputlisting{figs/code/Reskin-mcp.code}
        \end{minipage}%
    };
    \node[codebox, anchor=north east] (reskin-cloud) at (\linewidth, 0) {%
        \begin{minipage}{0.47\linewidth}
            \vspace*{2pt}
            \lstinputlisting{figs/code/Reskin-cloud.code}
        \end{minipage}%
    };

    \node[boxlabel] at ([xshift=5pt]reskin-mcp.north west)
        {MCP tool host {\bfseries\color{darkBlue}\{T\}}};
    \node[boxlabel] at ([xshift=5pt]reskin-cloud.north west)
        {webhost {\bfseries\color{darkBlue}\{T\}}};
\end{tikzpicture}
\endgroup
    \caption{The MCP host (left) executes the action associated with an unused
        user approval and delivers the result to the conversation through
        a custom callback \texttt{sendToChat}.
        The website host (right) calls the client's \texttt{archiveBuild}
        callback to save a copy of the built artifact before verification and
        publication; substituting \texttt{publish} for that callback publishes
        before verification.}
    \label{fig:reskin}
\end{figure}

\section{Complete Definitions}
\label{sec:ifc-static-semantics}

\subsection{Helper notations}
\label{sec:notation-overload}
We overload $\join$ and write $\tau\join\ell$ to join $\ell$ with
the label of $\tau$, and $\downarrow$ to
set the label of $\tau$ to $\ell$.
\[
  \begin{gathered}
    \funptrtype{\tau_i}{\tau_o}{\pcex}{\pctop}{\ell}\join\ell'
    = \funptrtype{\tau_i}{\tau_o}{\pcex}{\pctop}{\ell\join
      \ell'}
    \quad
    \tboolof{\ell}\join\ell' = \tboolof{\ell\join\ell'}
    \quad
    \tnatof{\ell}\join\ell' = \tnatof{\ell\join\ell'}
    \\
  \exprdowngrade{\tboolof{\ell'}}{\ell}=\tboolof{\ell}
  \qquad
  \exprdowngrade{\tnatof{\ell'}}{\ell}=\tnatof{\ell}
  \qquad
  \exprdowngrade{\tunit}{\ell}=\tunit.
  \qquad
  \exprdowngrade{\tempty}{\ell}=\tempty.
  \end{gathered}
\]

We also define
$\coreiface\asymp\coreiface'$, where
$\coreiface=(\sigtable,\celltable)$ and
$\coreiface'=(\sigtable',\celltable')$, to
mean the two interfaces
agree on every function and cell they both
declare.  That is,
\begin{align*}
  &\forall\,\funid{p}{f},s,s'.\;
    \sigtable(\funid{p}{f})=s\land\sigtable'(\funid{p}{f})=s'
    \implies s=s'
    \tag{functions agree}\\
  \land\;&\forall\,p,h,\tau,\tau'.\;
           \celltable(p,h)=\tau\land\celltable'(p,h)=\tau'
           \implies\tau=\tau'
           \tag{cells agree}
\end{align*}

We also define linking the signature
$\sigtable\bowtie\sigtable_{\calA}$ and cell table
$\celltable\bowtie\celltable_{\calA}$ as follows:
\begin{align*}
  (\sigtable\bowtie\sigtable_{\calA})(\funid{p}{f})&=
  \begin{cases}
    \sigtable(\funid{p}{f}) & \text{if }\funid{p}{f}\in\dom(\sigtable)\\
    \sigtable_{\calA}(\funid{p}{f}) & \text{otherwise}
  \end{cases}\\
  (\celltable\bowtie\celltable_{\calA})(p,h)&=
  \begin{cases}
    \celltable(p, h) & \text{if }(p,h)\in\dom(\sigtable)\\
    \celltable_{\calA}(p,h) & \text{otherwise}
  \end{cases}\\
\end{align*}

\subsection{IFC typing}
\Cref{fig:core-typing-full} gives the full IFC expression typing,
which, as stated in the main text, is mostly standard~\cite{zm01b,zm01}.
Besides the rules of \cref{fig:core-typing}, the figure contains the standard
rules for values, variables, operators, let bindings, conditionals and
subsumption.  \ruleof{TI-Abort} gives $\abort$ every well-formed type.
\ruleof{TI-If} checks each branch at $\pc\join\ell$, where $\ell$ is
the label of the guard, and joins $\ell$ into the result type, since the
result reveals which branch ran.
\begin{figure*}
  \centering
  \begin{judgments}
    \judgment{\ctyping{\Gamma}{\pc}{p}{\pcself{\text{max}}}{e}{\tau}}
  \end{judgments}
  \begin{mathpar}
    \inferrule*[Lab=TI-Unit]
      { }
      {\ctyping{\Gamma}{\pc}{p}{\pcself{\text{max}}}{\unit}{\tunit}}
    \and
    \inferrule*[Lab=TI-Bool]
      { }
      {\ctyping{\Gamma}{\pc}{p}{\pcself{\text{max}}}{b}{\tboolof{\ell}}}
    \and
    \inferrule*[Lab=TI-Nat]
      { }
      {\ctyping{\Gamma}{\pc}{p}{\pcself{\text{max}}}{n}{\tnatof{\ell}}}
    \and
    \inferrule*[Lab=TI-Var]
      {\Gamma(x)=\tau}
      {\ctyping{\Gamma}{\pc}{p}{\pcself{\text{max}}}{x}{\tau}}
    \and
    \inferrule*[Lab=TI-Fun]
      {\sigtable(\funid{q}{f})=
         \funsig{\tau_i}{\tau_o}{\pcex}{\pctop}
       \and q\flowsto\ell}
      {\ctyping{\Gamma}{\pc}{p}{\pcself{\text{max}}}
        {\funid{q}{f}}
        {\funptrtype{\tau_i}{\tau_o}{\pcex}{\pctop}{\ell}}}
    \and
    \inferrule*[Lab=TI-Abort]
      {\wfcore{\tau}}
      {\ctyping{\Gamma}{\pc}{p}{\pcself{\text{max}}}{\abort}{\tau}}
    \and
    \inferrule*[Lab=TI-Op]
      {\vdash\op:\tau_1\times\tau_2\longrightarrow\tau\\\\
       \ctyping{\Gamma}{\pc}{p}{\pcself{\text{max}}}{v_i}{\tau_i}
       \quad(i\in\{1,2\})}
      {\ctyping{\Gamma}{\pc}{p}{\pcself{\text{max}}}
        {\exprop{v_1}{v_2}}{\tau}}
    \and
    \inferrule*[Lab=TI-Downgrade]
      {\ctyping{\Gamma}{\pc}{p}{\pcself{\text{max}}}{v}{\tau}
       \and \exprdowngrade{\tau}{\ell}\text{ is defined}}
      {\ctyping{\Gamma}{\pc}{p}{\pcself{\text{max}}}
        {\exprdowngrade{v}{\ell}}{\exprdowngrade{\tau}{\ell}}}
    \and
    \inferrule*[Lab=TI-Let]
      {\ctyping{\Gamma}{\pc}{p}{\pcself{\text{max}}}{e_1}{\tau_1}
       \and
       \ctyping{\tau_1::\Gamma}{\pc}{p}{\pcself{\text{max}}}{e_2}{\tau_2}}
      {\ctyping{\Gamma}{\pc}{p}{\pcself{\text{max}}}
        {\exprlet{x}{e_1}{e_2}}{\tau_2}}
    \and
    \inferrule*[Lab=TI-Sub]
    {\wfcore{\tau}
      \and \tau'\subtypeof\tau
      \\\\
      \ctyping{\Gamma}{\pc}{p}{\pcself{\text{max}}}{e}{\tau'}
       \and 
       }
      {\ctyping{\Gamma}{\pc}{p}{\pcself{\text{max}}}{e}{\tau}}
    \and
    \inferrule*[Lab=TI-If]
    {
       \tau_i\subtypeof\tau
       \quad(i\in\{1,2\}) \and
       \wfcore{\tau} \\\\
      \ctyping{\Gamma}{\pc}{p}{\pcself{\text{max}}}{v}{\tboolof{\ell}} \and
       \ctyping{\Gamma}{\pc\join\ell}{p}{\pcself{\text{max}}}{e_i}{\tau_i}}
      {\ctyping{\Gamma}{\pc}{p}{\pcself{\text{max}}}
        {\exprif{v}{e_1}{e_2}}{\tau\join\ell}}
    \and
    \inferrule*[Lab=TI-Read]
      {\celltable(p,h)=\tau
       \and \pc\flowsto\labelof{\tau}}
      {\ctyping{\Gamma}{\pc}{p}{\pcself{\text{max}}}{\readexpr{h}}{\tau}}
    \and
    \inferrule*[Lab=TI-Write]
      {\celltable(p,h)=\tau \and \pc\flowsto\labelof{\tau}\\\\
       \ctyping{\Gamma}{\pc}{p}{\pcself{\text{max}}}{v}{\tau'}
       \and\tau'\subtypeof\tau
       }
      {\ctyping{\Gamma}{\pc}{p}{\pcself{\text{max}}}
        {\writeexpr{h}{v}}{\tunit}}
    \and
    \inferrule*[Lab=TI-App]
      {\ctyping{\Gamma}{\pc}{p}{\pcself{\text{max}}}{v_1}
         {\funptrtype{\tau_i}{\tau_o}{\pcex}{\pctop}{\ell}}
       \and
       \ctyping{\Gamma}{\pc}{p}{\pcself{\text{max}}}{v_2}{\tau}
       \and \tau\subtypeof\tau_i\\
       \pc\join\ell\flowsto\pcex
       \and \pcex\flowsto\pctop\join p
       \and \pcself{\text{max}}\flowsto\pctop}
      {\ctyping{\Gamma}{\pc}{p}{\pcself{\text{max}}}
        {\pcapp{v_1}{v_2}{\pcex}}{\tau_o\join\ell}}
  \end{mathpar}
  \caption{Full IFC expression typing.}
  \label{fig:core-typing-full}
\end{figure*}


An ordinary type $\sigma=\erase{\tau}$ drops every label and both pc
annotations of $\tau$,
\[
  \erase{\tboolof{\ell}}=\realbooltype\quad
  \erase{\tnatof{\ell}}=\realnattype\quad
  \erase{\tunit}=\tunit\quad
  \erase{\tempty}=\tempty\quad
  \erase{\funptrtype{\tau_i}{\tau_o}{\pcex}{\pctop}{\ell}}
    =\realfunctiontype{\erase{\tau_i}}{\erase{\tau_o}},
\]
and $\erase{\Gamma}$ replaces each $\tau$ in $\Gamma$ by
$\erase{\tau}$.  \Cref{fig:core-real-typing} gives the full
ordinary expression typing.  This typing is isomorphic to IFC typing
over the one-point label model, so each rule \ruleof{TR-X} is
\ruleof{TI-X} of \cref{fig:core-typing-full} with every label premise
removed and every type replaced by its erasure.  The mechanization
defines ordinary typing as this instance of IFC typing, which highlights
the correspondence between the two typings.
\begin{figure*}
  \centering
  \begin{judgments}
    \judgment{\rtyping{\Gamma}{p}{e}{\sigma}}
  \end{judgments}
  \begin{mathpar}[\mprset{lineskip=.8em,andskip=1.25em,sep=1em}]
    \inferrule*[Lab=TR-Unit]
      { }
      {\rtyping{\Gamma}{p}{\unit}{\tunit}}
    \and
    \inferrule*[Lab=TR-Bool]
      { }
      {\rtyping{\Gamma}{p}{b}{\realbooltype}}
    \and
    \inferrule*[Lab=TR-Nat]
      { }
      {\rtyping{\Gamma}{p}{n}{\realnattype}}
    \and
    \inferrule*[Lab=TR-Var]
      {\Gamma(x)=\sigma}
      {\rtyping{\Gamma}{p}{x}{\sigma}}
    \and
    \inferrule*[Lab=TR-Fun]
      {\sigtable(\funid{q}{f})=
         \funsig{\tau_i}{\tau_o}{\pcex}{\pctop}}
      {\rtyping{\Gamma}{p}{\funid{q}{f}}
        {\realfunctiontype{\erase{\tau_i}}{\erase{\tau_o}}}}
    \and
    \inferrule*[Lab=TR-Abort]
      { }
      {\rtyping{\Gamma}{p}{\abort}{\sigma}}
    \and
    \inferrule*[Lab=TR-Op]
      {\vdash\op:\sigma_1\times\sigma_2\longrightarrow\sigma \\\\
       \rtyping{\Gamma}{p}{v_i}{\sigma_i}
       \quad(i\in\{1,2\})}
      {\rtyping{\Gamma}{p}{\exprop{v_1}{v_2}}{\sigma}}
    \and
    \inferrule*[Lab=TR-Down]
      {\rtyping{\Gamma}{p}{v}{\sigma}
       \and \sigma\in\{\realbooltype,\realnattype,\tempty\}}
      {\rtyping{\Gamma}{p}{\exprdowngrade{v}{\ell}}
        {\sigma}}
    \and
    \inferrule*[Lab=TR-Let]
      {\rtyping{\Gamma}{p}{e_1}{\sigma_1}
       \and \rtyping{\sigma_1::\Gamma}{p}{e_2}{\sigma_2}}
      {\rtyping{\Gamma}{p}{\exprlet{x}{e_1}{e_2}}{\sigma_2}}
    \and
    \inferrule*[Lab=TR-Sub]
      {\rtyping{\Gamma}{p}{e}{\sigma'}
       \and \sigma'\subtypeof\sigma}
      {\rtyping{\Gamma}{p}{e}{\sigma}}
    \and
    \inferrule*[Lab=TR-If]
      {\rtyping{\Gamma}{p}{v}{\realbooltype}
       \and \rtyping{\Gamma}{p}{e_i}{\sigma_i}
       \and \sigma_i\subtypeof\sigma
       \quad(i\in\{1,2\})}
      {\rtyping{\Gamma}{p}{\exprif{v}{e_1}{e_2}}{\sigma}}
    \and
    \inferrule*[Lab=TR-Read]
      {\celltable(p,h)=\tau}
      {\rtyping{\Gamma}{p}{\readexpr{h}}{\erase{\tau}}}
    \and
    \inferrule*[Lab=TR-Write]
      {\celltable(p,h)=\tau
       \and \rtyping{\Gamma}{p}{v}{\sigma'}
       \and \sigma'\subtypeof\erase{\tau}}
      {\rtyping{\Gamma}{p}{\writeexpr{h}{v}}{\tunit}}
    \and
    \inferrule*[Lab=TR-App]
      {\rtyping{\Gamma}{p}{v_1}
         {\realfunctiontype{\sigma_i}{\sigma_o}}
       \and \rtyping{\Gamma}{p}{v_2}{\sigma}
       \and \sigma\subtypeof\sigma_i}
      {\rtyping{\Gamma}{p}{\pcapp{v_1}{v_2}{\pcex}}{\sigma_o}}
  \end{mathpar}
  \caption{Ordinary expression typing.}
  \label{fig:core-real-typing}
\end{figure*}

\subsection{World 3 semantics}
\label{sec:code-world-semantics}
Besides \ruleof{F-Code} and \ruleof{F-Code-Act}, World
$\mathrm W_3$ replaces the attacker cases of $\linkcall_{\mathrm W}$ and
$\linkret_{\mathrm W}$ in \cref{fig:core-linked-semantics} with the
following, and keeps all other rules of World $\mathrm W_2$.
\[
  \linkcall_{\mathrm W_3}(\store_{\calA},S,\pc,v_f,v)=
  \begin{cases}
    \langle\store_{\calA},\toattacker(\funid{q}{g},\subst{e}{x}{v})::S,\eventempty\rangle
      & \text{if } q\in\adv \text{ and } \bodytable_{\calA}(\funid{q}{g})=\lam{x}{e}\\
    \langle\store_{\calA},\toattacker(\funid{q}{g},X_{\calA}(\funid{q}{g})(v))::S,\eventempty\rangle
      & \text{if } q\in\adv \text{ and } X_{\calA}(\funid{q}{g}) \text{ is defined}\\
    \linkcall_{\mathrm W}(\store_{\calA},S,\pc,v_f,v)
      & \text{if } q\notin\adv\\
    \langle\store_{\calA},\emptyset,\eventabort\rangle & \text{otherwise}
  \end{cases}
\]
\[
  \linkret_{\mathrm W_3}(\store_{\calA},\phi,S,\funid{p}{f},v)=
  \begin{cases}
    \langle\store_{\calA},\toattacker(\funid{q}{g},E[v])::S,\trace_p\rangle
      & \text{if } \phi=\toattacker(\funid{q}{g},E[\pcapp{v_f}{u}{\ell}])\\
    \linkret_{\mathrm W}(\store_{\calA},\phi,S,\funid{p}{f},v) & \text{otherwise}
  \end{cases}
\]
where $v_f=\funid{q}{g}$ and $\trace_p$ is as in
\cref{fig:core-linked-semantics}.
To emphasize the correspondence among the worlds, the mechanization
adds an abstraction layer that generalizes the attacker's behavior.
This layer runs the heterogeneous system of the honest party and the
attacker on one unified semantics.  \ruleof{F-Code} and
\ruleof{F-Code-Act} instantiate the attacker abstraction with attacker
code, while \ruleof{F-Attacker} and \ruleof{F-Attacker-Act}
instantiate it with interactive state machines.

\section{Type Safety}
\label{sec:type-safety}
As a sanity check, we show that a well-typed program linked with a
well-typed (either IFC-typed or ordinarily typed) attacker never gets stuck.  The proof lifts the standard
progress and preservation lemmas for expressions to linked
configurations.  Both lemmas hold for IFC typing over any label
lattice, so they also hold for ordinary typing, which is IFC typing
over the one-point label model.

\subsection{World 2}
Progress states that a closed, IFC-typed expression never gets stuck.
A value steps to a return action by \ruleof{S-Ret}, and an expression
that reaches $\abort$ steps to an abort action by \ruleof{S-Abort}.
\begin{lemma}[IFC Progress]
  \label{lem:ideal-progress}
  Consider a principal $p$, labels $\pc$ and $\pctop$, an expression
  $e$ and a type $\tau$ such that
  $\ctyping{\emptyset}{\pc}{p}{\pctop}{e}{\tau}$.  Then for every store
  $\store$,
  \[
    \exists N,\store',\trace.\;
    \langle e,\store\rangle\xrightarrow{p}\langle N,\store',\trace\rangle
  \]
\end{lemma}

Preservation states that an internal step keeps the expression that
$\funid{p}{f}$ runs at the result type of $\funid{p}{f}$ and keeps the
cells of every principal well typed.  We write $\celltable|_q$ for the
restriction of $\celltable$ to the cells of principal $q$.
\begin{lemma}[IFC Preservation]
  \label{lem:ideal-preservation}
  Consider an interface $\coreiface=(\sigtable,\celltable)$, a function
  $\funid{p}{f}$, an expression $e$ and a store $\store$ such that
  $\vdash_{\mathrm W_2}\coreiface$, $\celltable|_p\vdash\store$ and
  $\ctyping{\emptyset}{p}{p}{\pctop}{e}{\tau_o\join p}$.  Then for every
  expression $e'$, store $\store'$ and trace $\trace$,
  \[
    \begin{aligned}
      \langle e,\store\rangle\xrightarrow{p}
        \langle\internaloutcome{e'},\store',\trace\rangle
      \implies
      (\ctyping{\emptyset}{p}{p}{\pctop}{e'}{\tau_o\join p})
      \;\land\;&(\forall q.\;
        \celltable|_q\vdash\store
        \implies
        \celltable|_q\vdash\store')
    \end{aligned}
  \]
  where $\sigtable(\funid{p}{f})=\funsig{\tau_i}{\tau_o}{\pcex}{\pctop}$.
\end{lemma}

Type safety lifts both lemmas to linked runs.  Every configuration
that a run reaches has well-typed honest cells and a well-typed running
honest frame, and it either has halted or takes another step.
\begin{theorem}[Type Safety in World 2]
  \label{thm:ideal-type-safety}
  Consider a program $\calP$, an attacker $\calA$ and a store
  $\store_{\mathrm{init}}$ such that $\vdash_{\mathrm W_2}\calP$,
  $\vdash_{\mathrm W_2}\calA$ and $\vdash\store_{\mathrm{init}}$.  Then for every
  configuration $C=\langle s_{\calA},\store,S\rangle$ and trace
  $\trace$,
  \[
    \begin{aligned}
      C_{\mathrm{init}}\longrightarrow^{*}_{\mathrm W_2}\langle C,\trace\rangle
      \implies
      &\vdash\store\\
      \land\;&\forall\funid{p}{f},e,S'.\;
        S=\tohonest(\funid{p}{f},e)::S'
        \implies
        \ctyping{\emptyset}{p}{p}{\pctop}{e}{\tau_o\join p}\\
      \land\;&(S=\emptyset
        \lor
        \exists C',\trace'.\;C\linkstep_{\mathrm W_2}\langle C',\trace'\rangle)
    \end{aligned}
  \]
  where
  $(\sigtable\bowtie\sigtable_{\calA})(\funid{p}{f})
   =\funsig{\tau_i}{\tau_o}{\pcex}{\pctop}$.
\end{theorem}

\subsection{World 1}
World 1 satisfies the same three results with ordinary typing in
place of IFC typing.  In this subsection, the store typing $\vdash$
checks each cell against its ordinary type with ordinary typing.  Ordinary
preservation needs no well-formed interface, since every erased
interface is well formed.
\begin{lemma}[Ordinary Progress]
  \label{lem:real-progress}
  Consider a principal $p$, an expression $e$ and an ordinary type $\sigma$ such
  that $\rtyping{\emptyset}{p}{e}{\sigma}$.  Then for every
  store $\store$,
  \[
    \exists N,\store',\trace.\;
    \langle e,\store\rangle\xrightarrow{p}\langle N,\store',\trace\rangle
  \]
\end{lemma}
\begin{lemma}[Ordinary Preservation]
  \label{lem:real-preservation}
  Consider an interface $\coreiface=(\sigtable,\celltable)$, a function
  $\funid{p}{f}$, an expression $e$ and a store $\store$ such that
  $\celltable|_p\vdash\store$ and
  $\rtyping{\emptyset}{p}{e}{\erase{\tau_o}}$.  Then for every
  expression $e'$, store $\store'$ and trace $\trace$,
  \[
    \begin{aligned}
      \langle e,\store\rangle\xrightarrow{p}
        \langle\internaloutcome{e'},\store',\trace\rangle
      \implies
      (\rtyping{\emptyset}{p}{e'}{\erase{\tau_o}})
      \;\land\;&(\forall q.\;
        \celltable|_q\vdash\store
        \implies
        \celltable|_q\vdash\store')
    \end{aligned}
  \]
  where $\sigtable(\funid{p}{f})=\funsig{\tau_i}{\tau_o}{\pcex}{\pctop}$.
\end{lemma}
\begin{theorem}[Type Safety in World 1]
  \label{thm:real-type-safety}
  Consider a program $\calP$, an attacker $\calA$ and a store
  $\store_{\mathrm{init}}$ such that $\vdash_{\mathrm W_1}\calP$,
  $\vdash_{\mathrm W_1}\calA$ and $\vdash\store_{\mathrm{init}}$.  Then for every
  configuration $C=\langle s_{\calA},\store,S\rangle$ and trace
  $\trace$,
  \[
    \begin{aligned}
      C_{\mathrm{init}}\longrightarrow^{*}_{\mathrm W_1}\langle C,\trace\rangle
      \implies
      &\vdash\store\\
      \land\;&\forall\funid{p}{f},e,S'.\;
        S=\tohonest(\funid{p}{f},e)::S'
        \implies
        \rtyping{\emptyset}{p}{e}{\erase{\tau_o}}\\
      \land\;&(S=\emptyset
        \lor
        \exists C',\trace'.\;C\linkstep_{\mathrm W_1}\langle C',\trace'\rangle)
    \end{aligned}
  \]
  where
  $(\sigtable\bowtie\sigtable_{\calA})(\funid{p}{f})
   =\funsig{\tau_i}{\tau_o}{\pcex}{\pctop}$.
\end{theorem}

\subsection{World 3}
Type safety in world $\mathrm W_3$ is syntactic.  The judgment
$\vdash_{\mathrm W_3}\calA$ types the attacker code itself rather than
the actions that the attacker emits, so the invariant also types the
running attacker frame and the attacker store.  The attacker code is
typed in world $\mathrm W_2$, so the invariant uses IFC typing.
\begin{theorem}[Type Safety in World 3]
  \label{thm:code-type-safety}
  Consider a program $\calP$, attacker code $\calA$ and a store
  $\store_{\mathrm{init}}$ such that $\vdash_{\mathrm W_2}\calP$,
  $\vdash_{\mathrm W_3}\calA$ and $\vdash\store_{\mathrm{init}}$.  Then for every
  configuration $C=\langle\store_{\calA},\store,S\rangle$ and trace
  $\trace$,
  \[
    \begin{aligned}
      C_{\mathrm{init}}\longrightarrow^{*}_{\mathrm W_3}\langle C,\trace\rangle
      \implies
      &\vdash\store
      \land\celltable_{\calA}\vdash\store_{\calA}\\
      \land\;&\forall\funid{p}{f},e,S'.\;
        \left(
          S=\tohonest(\funid{p}{f},e)::S'
          \lor
          S=\toattacker(\funid{p}{f},e)::S'
        \right)\\
      &\quad\implies
        \ctyping{\emptyset}{p}{p}{\pctop}{e}{\tau_o\join p}\\
      \land\;&(S=\emptyset
        \lor
        \exists C',\trace'.\;C\linkstep_{\mathrm W_3}\langle C',\trace'\rangle)
    \end{aligned}
  \]
  where
  $(\sigtable\bowtie\sigtable_{\calA})(\funid{p}{f})
   =\funsig{\tau_i}{\tau_o}{\pcex}{\pctop}$.
\end{theorem}

Type safety in world $\mathrm W_3$ also holds in the traditional form of
initialization, preservation and progress.  This form uses a typing
judgment on configurations that also types the suspended frames.
\begin{definition}[Configuration Typing]
  \label{def:code-config-typing}
  Consider a program $\calP$ and attacker code $\calA$.  A configuration
  $C=\langle\store_{\calA},\store,S\rangle$ is well typed, written
  $\vdash_{\mathrm W_3}C$, if $\vdash\store$,
  $\celltable_{\calA}\vdash\store_{\calA}$ and either $S=\emptyset$ or
  $S=\phi::S'$ with
  \begin{align*}
    &(\phi=\tohonest(\funid{p}{f},e)\land p\notin\adv
      \lor\phi=\toattacker(\funid{p}{f},e)\land p\in\adv)
      \tag{where $p$ belongs}\\
    \land\;&\ctyping{\emptyset}{p}{p}{\pctop}{e}{\tau_o\join p}
      \tag{running frame is well-typed}\\
    \land\;&\funid{p}{f}\vdash S'
      \tag{suspended frames are well-typed}
  \end{align*}
  The judgment $\funid{q}{g}\vdash S$ states that each frame of $S$
  resumes well typed on a well-typed result of the frame above it.  It
  holds for $S=\emptyset$ and for $S=\phi::S'$ with
  \begin{align*}
    &(\phi=\tohonest(\funid{p}{f},E[\pcapp{v_f}{u}{\ell}])\land p\notin\adv
      \lor\phi=\toattacker(\funid{p}{f},E[\pcapp{v_f}{u}{\ell}])\land p\in\adv)
      \tag{where $p$ belongs}\\
    \land\;&\forall v.\;
      \ctyping{\emptyset}{q}{q}{\pctop'}{v}{\tau_o'\join q}
      \implies
      \ctyping{\emptyset}{p}{p}{\pctop}{E[v]}{\tau_o\join p}
      \tag{resumption is well-typed}\\
    \land\;&\funid{p}{f}\vdash S'
      \tag{suspended frames are well-typed}
  \end{align*}
  where
  $(\sigtable\bowtie\sigtable_{\calA})(\funid{p}{f})
   =\funsig{\tau_i}{\tau_o}{\pcex}{\pctop}$ and
  $(\sigtable\bowtie\sigtable_{\calA})(\funid{q}{g})
   =\funsig{\tau_i'}{\tau_o'}{\pcex'}{\pctop'}$.
\end{definition}

\begin{theorem}[Configuration Safety in World 3]
  \label{thm:code-config-safety}
  Consider a program $\calP$, attacker code $\calA$ and a store
  $\store_{\mathrm{init}}$ such that $\vdash_{\mathrm W_2}\calP$,
  $\vdash_{\mathrm W_3}\calA$ and $\vdash\store_{\mathrm{init}}$.  Then
  $\vdash_{\mathrm W_3}C_{\mathrm{init}}$, and for all configurations $C$ and $C'$ and
  traces $\trace$,
  \begin{align*}
    &\vdash_{\mathrm W_3}C
      \land C\linkstep_{\mathrm W_3}\langle C',\trace\rangle
      \implies\vdash_{\mathrm W_3}C'
      \tag{preservation}\\
    &C_{\mathrm{init}}\longrightarrow^{*}_{\mathrm W_3}\langle C,\trace\rangle
      \implies S=\emptyset
      \lor\exists C'',\trace'.\;
        C\linkstep_{\mathrm W_3}\langle C'',\trace'\rangle
      \tag{progress}
  \end{align*}
  where $C=\langle\store_{\calA},\store,S\rangle$.
\end{theorem}

Theorem~\ref{thm:code-config-safety} implies
Theorem~\ref{thm:code-type-safety}.  Initialization and preservation
give $\vdash_{\mathrm W_3}C$ for every reachable configuration $C$, and
$\vdash_{\mathrm W_3}C$ contains the store and running-frame conditions
of Theorem~\ref{thm:code-type-safety}.

\end{document}